\documentclass[prc,twocolumn,secnumroman,tightenlines,amssymb,aps,nobibnotes,superscriptaddress,balancelastpage,floatfix,nofootinbib]{revtex4}
\usepackage{soul}  

\usepackage{amsmath}
\usepackage{slashbox}
\usepackage{soul}
\usepackage{float}
\usepackage{lipsum}
\usepackage{graphicx}
\usepackage{multirow}           
\usepackage{color}
\DeclareRobustCommand\substyle{\name@idx{document substyle}}
\DeclareRobustCommand\classoption{\name@idx{document class option}}
\DeclareRobustCommand\classname{\name@idx{document class}}
\def\name@idx#1#2{{\ttfamily#2}
\index{#2\space#1=\string\ttt{#2}\space#1}\index{#1>#2=\string\ttt{#2}}}

\DeclareTextFontCommand{\rb}{\color{red}\bfseries}

\begin{document}







\title{Neutron emission from the photon-induced reactions
in ultraperipheral ultrarelativistic heavy-ion collisions}

\author{P. Jucha}
\email{Pawel.Jucha@ifj.edu.pl}
\affiliation{Institute of Nuclear Physics PAN, ul.\,Radzikowskiego 152,Pl-31342 Krak\'ow, Poland}

\author{M. K\l{}usek-Gawenda}
\email{Mariola.Klusek-Gawenda@ifj.edu.pl }
\affiliation{Institute of Nuclear Physics PAN, ul.\,Radzikowskiego 152,Pl-31342 Krak\'ow, Poland}  

\author{A. Szczurek}
\email{Antoni.Szczurek@ifj.edu.pl}
\affiliation{Institute of Nuclear Physics PAN, ul.\,Radzikowskiego 152,Pl-31342 Krak\'ow, Poland}
\affiliation{College of Natural Sciences, Institute of Physics,
            University of Rzesz\'ow, ul. Pigonia 1, PL-35-310 Rzesz\'ow, Poland}
              
\author{M. Ciema\l{}a}
\email{Michal.Ciemala@ifj.edu.pl}
\affiliation{Institute of Nuclear Physics PAN, ul.\,Radzikowskiego 152,Pl-31342 Krak\'ow, Poland}
           
\author{K. Mazurek}
\email{Katarzyna.Mazurek@ifj.edu.pl}
\affiliation{Institute of Nuclear Physics PAN, ul.\,Radzikowskiego 152,Pl-31342 Krak\'ow, Poland}

\date{\today}

\begin{abstract}
The ultraperipheral collisions are the source of various interesting phenomena based on photon-induced reactions. 
We calculate cross sections for single and any number of n, p, $\alpha$, $\gamma$-rays in ultraperipheral heavy-ion collision for LHC energies. 

We analyze the production of a given number of neutrons
relevant for a recent ALICE experiment, for $\sqrt{s_{NN}}$ = 5.02~TeV.
In our approach, we include both single and multiple photon exchanges as well as
the fact that not all photon energies are used in the process
of equilibration of the residual nucleus. 
We propose a simple two-component model in which only part of photon energy $E_\gamma$ is changed into the excitation energy of the nucleus ($E_{exc} \neq E_{\gamma}$) and compare its results with outcomes of HIPSE and EMPIRE codes. The role of high photon energies for small neutron multiplicities is discussed. Emission of a small number of neutrons at high photon energies seems to be crucial to understand the new ALICE data. All effects work in the desired direction, but the description of the cross section of four- and five-neutron emission cross sections from first principles is rather demanding. The estimated emission of charged particles such as protons, deuterons and $\alpha$ is shortly discussed and confronted with very recent ALICE data, obtained with the proton Zero Degree Calorimeter.

\end{abstract}


%


\maketitle


\section{Introduction}
\label{Sect.I}
The photon-induced reactions are a wonderful playground for investigation of various phenomena, starting from Giant Dipole Resonance (GDR) by particle emission from excited nuclei and even fission of the excited nucleus up to the production of high-energy particles such as Deltas and other nucleon resonances.

 The collision of the two lead nuclei with ultrarelativistic velocities
 are the subject of investigation of many groups associated with RHIC ~\cite{Chiu:2001ij} and the
 LHC~\cite{ALICE502a,ALICE_protons}. Depending on the impact parameters, for central collision, the
 quark-gluon plasma physics is crucial, but for ultraperipheral
 collisions (UPC), other, more nuclear physics type effects start to 
play the dominant role. 

The exchange of the photon between colliding nuclei leads to electromagnetic disintegration of the nuclei. The Coulomb dissociation is an important ingredient influencing the lifetime of nuclear beams at RHIC and the LHC \cite{cs5}.

The main physics idea behind the appearance of such processes is quite simple: the motion of charged
nuclei is the source of moving electromagnetic (EM) fields. The quasi-real photons created by the EM field of one nucleus, can hit the second nucleus and excite
it. The standard Coulomb interaction of two charged nuclei with 
$Z_1=Z_2$ = 82 predicts the distribution of the excitation energy 
of such spectators up to about 50~MeV. Thus, our previous calculations presented in Ref.~\cite{Klusek-Gawenda:2013ema} have been restricted to this value. However, the new measurements 
of the ALICE collaboration \cite{ALICE502a} suggest 
that there are other phenomena which should be taken into 
consideration when the nucleus excitation energy achieves higher values. 
In the present study, we discuss a range of a photon energy relevant for production of a few neutrons.

In the UPC the excitation of a nucleus is obtained naively with a convolution of the photon flux and the photoabsorption cross section. Detailed calculations requires precise modeling of the energy transfer to the nucleus and estimation of the neutron emission from the nucleus, both in preequilibrium and equilibrium stage. The cross section for neutron emission can be also obtained by similar convolution of the photon flux with the photoneutron cross section. The latter was measured up to 140~MeV photon energies \cite{LEPRETRE1978}. 

It is very difficult to measure precisely neutrons for large photon energies in $\gamma+A$ collisions. However, it is easier to detect neutrons in UPC of heavy ions due to the narrow emission cone.
The ALICE experiment provides neutron multiplicities, but other particles, such as protons and alpha particles, were not measured and discussed so far.

In Ref.~\cite{Klusek-Gawenda:2013ema} we assumed that the full photon energy is transformed into excitation energy of the nucleus ($E_{exc} = E_{\gamma}$)
and the GEMINI++ program \cite{gemini} was used to deexcite the hot nucleus, assuming Hauser-Feshbach \cite{hauser} particle cascade model. Furthermore we compared our results with 
the production of multiple neutrons (up to 3 neutrons) 
as measured by the ALICE collaboration in UPC for $\sqrt{s_{NN}}$ = 2.76~TeV, see Ref.~\cite{ALICE:2012aa}. A new valuable result was obtained recently
by the ALICE collaboration for $\sqrt{s_{NN}}$ = 5.02~TeV \cite{ALICE502a}, where up to 5 neutrons were measured.
As it will be discussed below, our old approach leads to a smaller number of neutron emissions than measured by the ALICE collaboration.  

In this paper we shall discuss how the new ALICE result is related to the underlying physics. 
The presented method consists of three stages: 

\noindent (1) the calculation of the photon flux within the equivalent photon approximation (EPA);

\noindent (2) absorption of photons by nucleus and excitation energy estimation using: a simple two-component model (TCM),  Heavy Ion Phase -- Space Exploration (HIPSE) \cite{hipse1} or Nuclear Reaction Model Code System
for Data Evaluation (EMPIRE) \cite{empire}; 

\noindent (3) deexctitation of the hot nucleus in GEMINI++ statistical code.

\noindent Then the results are compared with experimental data. The novelty is a proposition of a simple two-component model (TCM) where $E_{exc} < E_{\gamma}$, consistent with $\gamma+A\rightarrow A' + kn$ experimental data for $E_{\gamma}<140$~MeV. The role of high-energy photon scattering off nuclei will be emphasized.

\section{Theoretical methods}
\label{Sect.II}
\subsection{Photon flux}

The moving charged nuclei generate the photon flux, which is
calculated using the equivalent photon approximation (EPA) \cite{b-space-EPA}. 
Analytic formula for point-like nucleus can be found e.g. in~\cite{Jackson}:

\begin{equation}
N(\omega, b) = \frac{\alpha Z^2}{\pi^2} \frac{u^2}{\beta^2 \omega b^2} \left( K^2_1(u) + \frac{1}{\gamma_{LAB}^2} K^2_0(u) \right).  
\label{Eq.1}
\end{equation}
For extended charge, one has:

\begin{equation}
            N(\omega,b) = \frac{Z^2 \alpha_{em}}{\pi^2 \beta^2}\frac{1}{\omega b^2} \times \left|
            \int d\chi \hspace{3pt} \chi^2 
            \frac{F(\frac{\chi^2+u^2}{b^2})}{\chi^2+u^2}J_1(\chi) \right|^2 .
\label{Eq.2}
\end{equation}

\noindent Above $\omega$ is photon energy and $b$ is transverse distance of photons from the emitting nucleus. If we are interested in photon energy in the rest frame of the absorbing nucleus: $\omega=E_{\gamma}$. The electromagnetic form factor of the nucleus $F(\frac{\chi^2+u^2}{b^2})$ depends on $u = \frac{\omega b}{\gamma_{CM}\beta}$, $\gamma_{CM}=\frac{1+\beta^2}{1-\beta^2}$, $\chi$ is an auxiliary dimensionless variable related to photon transverse momentum via the relation $\chi = k_\perp b$. In realistic case $F(\frac{\chi^2+u^2}{b^2})$ is calculated as a Fourier transform of the nucleus charge density.

\begin{figure}[h]
    \centering
        a)\includegraphics[width=0.47\textwidth]{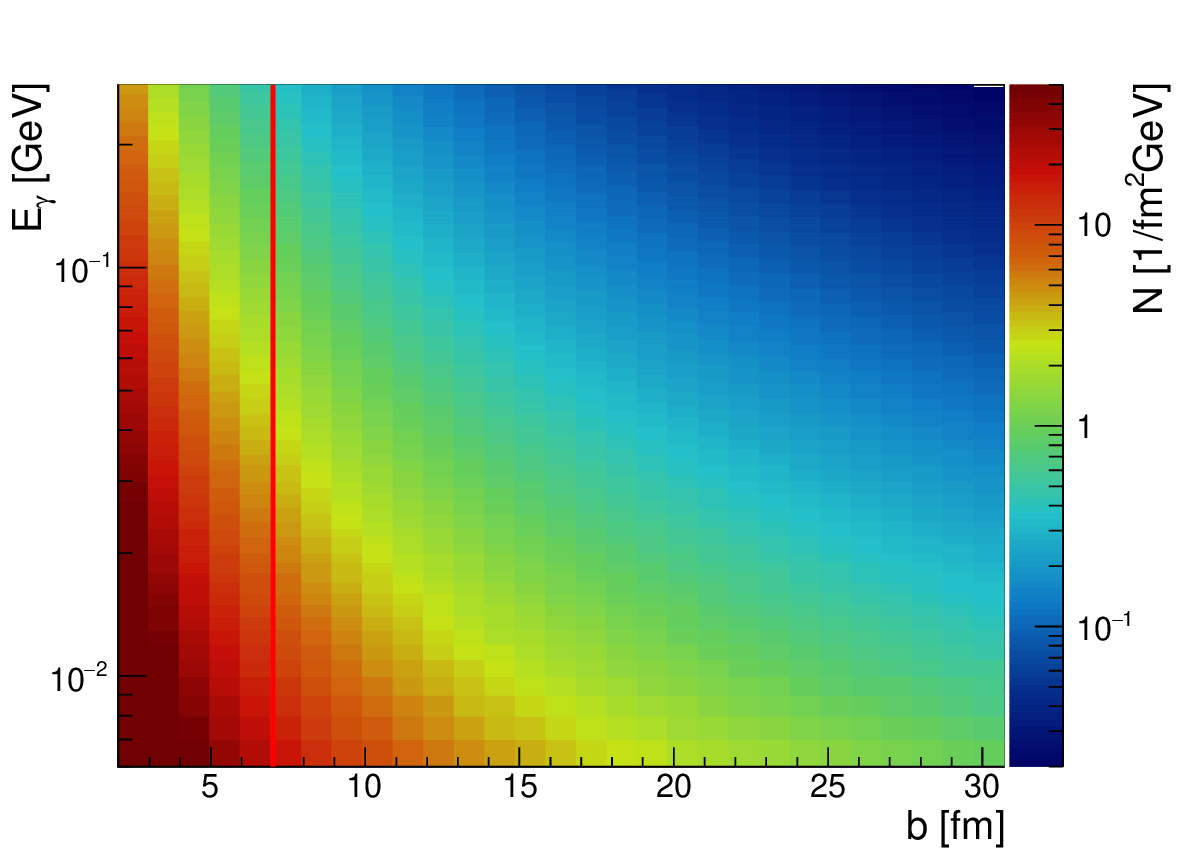}
        b)\includegraphics[width=0.47\textwidth]{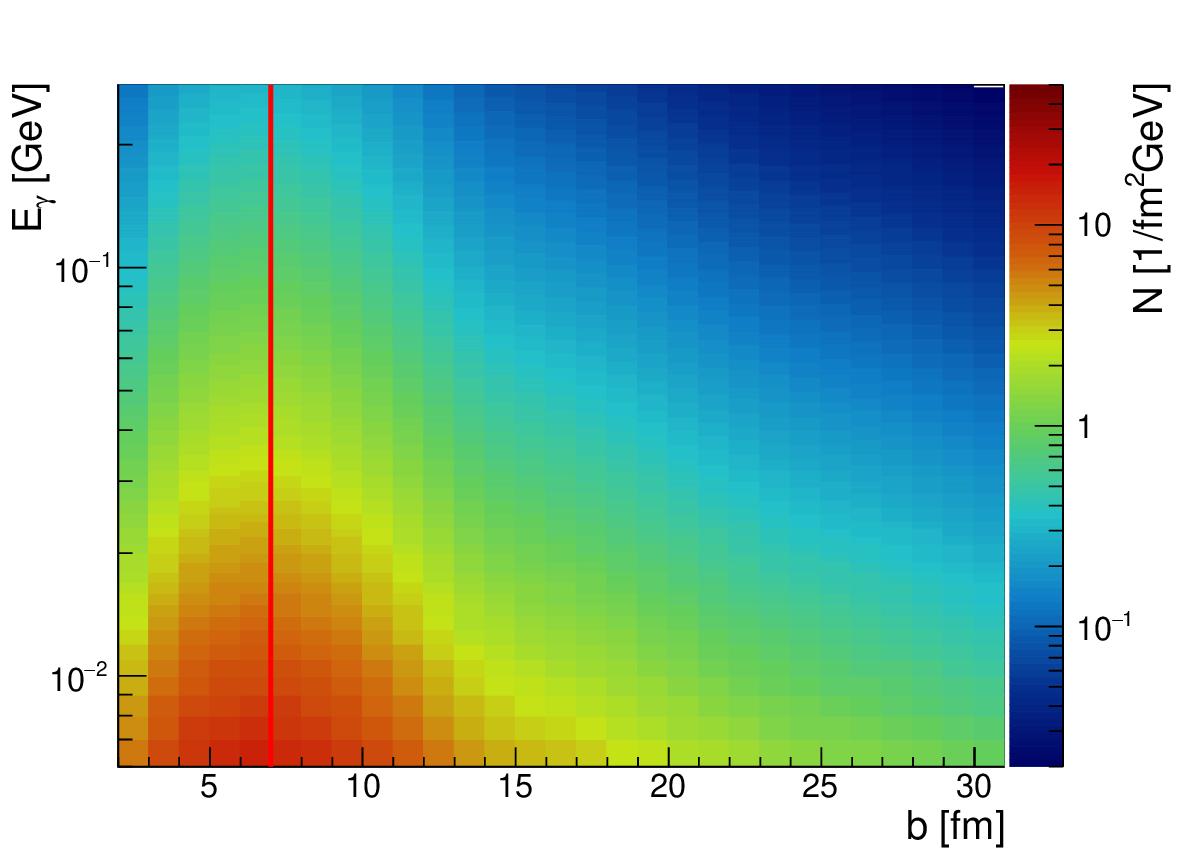}
    \caption{Calculated photon flux dependence on photon energy and impact parameter.  The $b_1=R_A$ impact parameter is marked explicitly by the red line. a) Analytic formula (Eq.~(\ref{Eq.1})); b) Flux obtained with realistic form factor (Eq.~(\ref{Eq.2})). Here, the Fermi functional form of charge density was used to calculate the form factor.  }
    \label{fig:flux}
\end{figure}

The EPA method allows us to calculate the impact parameter space distribution of photons produced by fast moving charged sources. In the Eq.~(\ref{Eq.2}), $N(\omega,b)$ means rather $\frac{dN(\omega,b)}{d\omega d^2b}$ but is used to shorten other more complicated formulas. The photon flux distribution for ultraperipheral collisions of  $^{208}$Pb nuclei with the energy $\sqrt{s_{NN}}$ = 5.02~TeV are shown in Fig.~\ref{fig:flux}. Photons are produced typically with energies below 20~MeV, but there is still a sizable probability to generate highly energetic $\gamma$-rays.
This energetic quasi-real photons could appear for the impact parameter $b = R_{A_1} + R_{A_2}>$14~fm, where UPC is considered.
The nucleus absorbs such photons, raising its internal energy, and follows typical nuclear processes such as $\gamma$-rays emission and particle evaporation, fission, or even multifragmentation.
\begin{figure}[!bt]
\centering
\includegraphics[width=6cm]{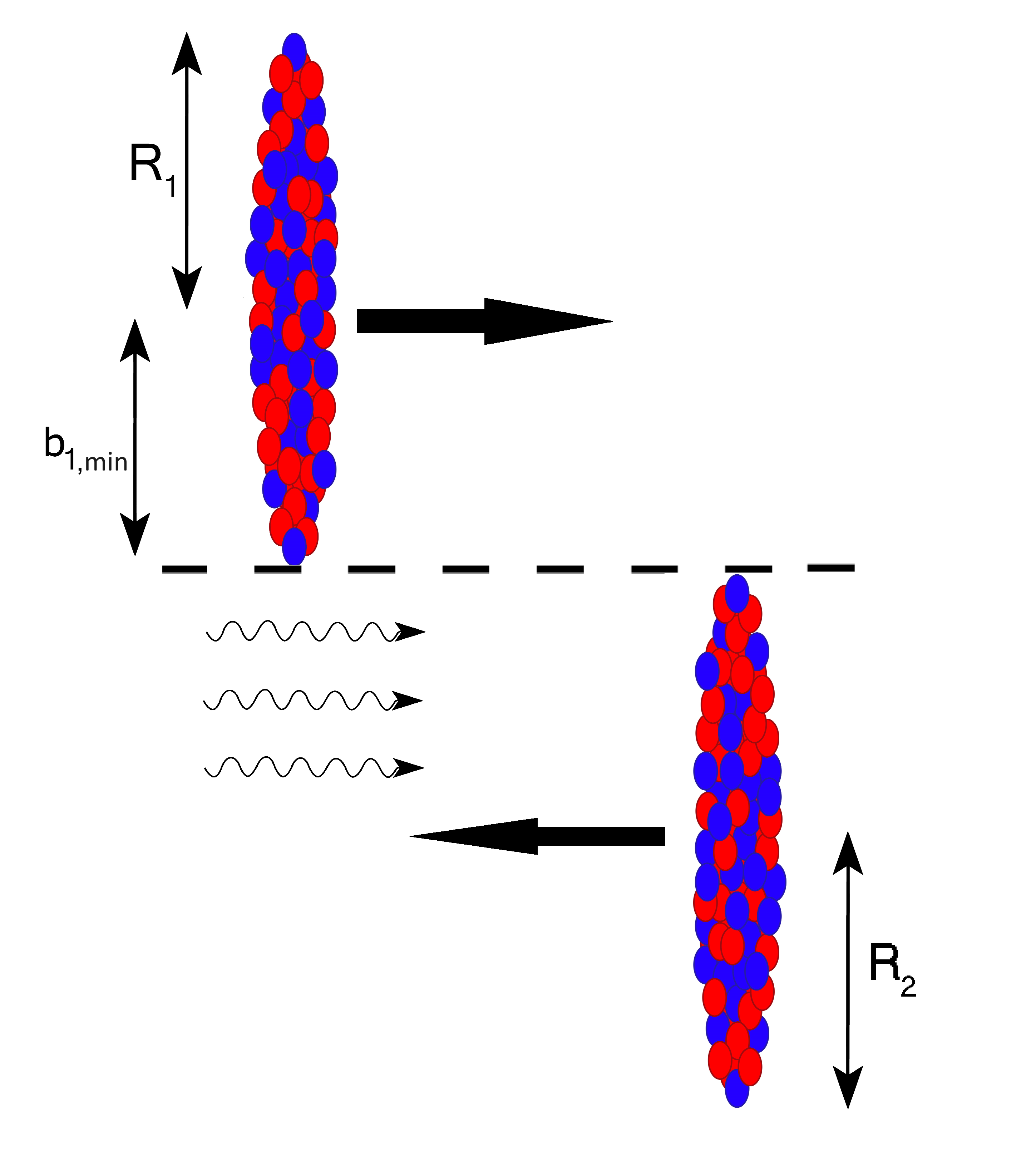}
\caption{  The impact parameter range of quasi-real photons emitted by one nucleus which excite the second nucleus.}
\label{fig:integration_over_d2b}
\end{figure}

The flux of photons depends on the transverse distance from the emitting nucleus, where quasi-real
photons are emitted.
This is illustrated in Fig.~\ref{fig:integration_over_d2b}. 
In our calculations, one has to integrate over all configurations of the impact parameter space
when quasi-real photon associated with one nucleus hits
the second nucleus.
The impact parameter $b_1 = R_A$ configuration is a limiting
configuration which leads to the excitation of the second nucleus.
In the present calculations, the impact parameter $b_1$ is integrated in the range 7~--~10$^9$~fm.

\subsection{Existing generators for photon nucleus interacion}

A first dedicated $\gamma + A$ generator was presented in
Ref.~\cite{C94}. This generator was valid up to $E_{\gamma}$ = 1~GeV.
The RELDIS code \cite{reldis} seems to be currently the best photon-induced 
intranuclear cascade code, at least up to a few GeV photon energy
as it takes into account a multitude of hadronic final states
as well as corresponding secondary interactions\footnote{The RELDIS code is not accessible to us.}. There the integration is done over the
transverse distance between colliding nuclei.

In Ref.~\cite{IPBSMMR1997} many multi-body channels were discussed
and Monte Carlo calculations confronted with existing 
experimental data for a few body final states.
For multipion final states (3~$\pi$~--~8~$\pi$) an approximate 
statistical approaches were considered and discussed. 
The consequences for subsequent neutron emission in the process 
of residual nucleus deexcitation were not discussed in 
\cite{IPBSMMR1997}.
The results of the study in \cite{IPBSMMR1997} were implemented
into the RELDIS code. We do not know about any detailed experimental study of 
neutron emission induced by high-energy photons.

At higher photon energies ($E_{\gamma} >$ 2~--~5~GeV) the physics again changes
and photon interaction directly with partons in protons and neutrons
must be included.
 
The intranuclear model in \cite{barashenkov} tried to describe neutron, proton and pion emission at energies below the two-pion production threshold. 
The Monte Carlo generator based on the dual parton model (DPMJET-III) \cite{Ranft:1997pu,dmp} is
capable of simulating hadron-hadron, hadron-nucleus, nucleus-nucleus, photon-hadron, photon-photon and photon-nucleus interactions from a few GeV up to the highest cosmic ray energies. 

The STARlight generator \cite{starlight}, often used for ultraperipheral collisions, can compute cross sections and generate events for two-photon and photonuclear interactions accompanied by mutual Coulomb excitation. These processes involve three or four photons: one or two for the photonuclear or two-photon interaction, plus one to excite each nucleus. The program can also calculate cross sections for reactions involving mutual excitation to a Giant Dipole Resonance, which typically decays through the emission of a single neutron. STARlight gives only fractions of the 0n0n, 1n1n, Xn0n, and XnXn categories. In \cite{NooN}, a very useful UPC Monte Carlo code \textbf{n}$^\mathbf{O}_\mathbf{O}$\textbf{n} is described which parametrizes rather existing experimental data for $\gamma+A \to A'+kn$.

Neutron production from excited nuclear system will be discussed
in future also at the EIC. A BeAGLE Monte Carlo generator has already been constructed. It allows to calculate average neutron multiplicity
in $e + A$ collisions \cite{BEAGLE}. BeAGLE uses intranuclear cascade
DPMJet and PYTHIA-6 for hadronization. The geometry in BeAGLE 
is, however, somewhat different (virtual photons) than the geometry 
needed in UPC (quasi-real photons).

A color dipole model would be the best
starting point in this region of energy.
However, a future approach would require the calculation of the energy
transfer to the nucleus due to the interaction of the color dipole
with the nuclear matter and modeling of the subsequent neutron
emissions. Again a very challenging task. Only emission of dijet
and pions was considered \cite{RER1996} within the dual photon approach.

\subsection{Towards simple dynamical model}
\label{Sect.II}

Now, we wish to consider the UPC of lead nuclei.
Starting from the one-photon excitation mechanism, the total cross
section depends on the flux of photons in a nucleus and photon 
absorptive cross section \cite{Klusek-Gawenda:2013ema}:
\begin{equation}
    \sigma\left(PbPb \xrightarrow[]{1\gamma} PbPb^*\right) = \int \int \sigma_{abs}\left( E_\gamma \right) N\left(E_\gamma,b\right) d^2b dE_\gamma .
\label{eq:sigma_tot_neutron_sum}    
\end{equation}
Here $N(E_\gamma,b)$ is photon flux associated with ultrarelativistic nucleus (see e.g. \cite{b-space-EPA} and Fig.~\ref{fig:flux}).
Rather, a broad range of energy and impact parameter are necessary to calculate the integral precisely. Please note that for large $b$ one has $\vec{b} \approx \vec{b_1}$ or $\vec{b} \approx \vec{b_2}$ for single nucleus excitation. Here, by $\vec{b_1}$ or $\vec{b_2}$, we understand the transverse distance of the photon (interaction point) from nucleus $1$ or nucleus $2$, respectively, so-called photon impact parameter. 

\begin{table*}[!bt]
\caption{Cross section $\sigma_{tot}$ [b] for generating a given number of neutrons in UPC for $\sqrt{s_{NN}} = 5.02$~TeV, for different values of $E_0$ parameter of the step-like TCM confronted with pure GEMINI++ and TCM with sin-like function predictions. In bold we show results for $E_0$~=~50~MeV. This value will be used throughout this paper.}
\begin{ruledtabular}
\begin{tabular}{c|c|c|c|c|ccccc}
& \multicolumn{8}{c}{$\sigma$ [b]} \\ \hline
& ALICE \cite{ALICE502a} &  RELDIS & GEMINI++ & \multicolumn{6}{c}{TCM}  \\ \hline
&   experiment &&& sin-like & \multicolumn{5}{c}{step-like}  \\ \hline
\backslashbox{kn}{E$_0$ [MeV]} &  & & & 50 & 30 & 40 & \bf{50} & 60 & 70       \\ \hline
\quad 1  \quad  &    108.4$\pm$3.90&\hfill 104.1$\pm$ 5.2  &\hfill 93.71 &\hfill 83.37&\hfill 99.39 &\hfill 99.06 &\hfill \bf{98.75} &\hfill 98.47 &\hfill 98.22  \\
\quad2    &          25.0$\pm$1.30 &\hfill 21.9$\pm$ 1.1 &\hfill25.06 &\hfill 21.16&\hfill 23.51 &\hfill 24.77 &\hfill \bf{25.55} &\hfill 26.05 &\hfill 26.38  \\
\quad3    &          7.95$\pm$0.25 &\hfill 7.59$\pm$ 0.38 &\hfill3.05 &\hfill  4.05 &\hfill 5.91  &\hfill 6.02  & \hfill\bf{6.07}  &\hfill 6.06  &\hfill 6.02   \\
\quad4    &          5.65$\pm$0.33 &\hfill4.29$\pm$ 0.22 &\hfill2.32 &\hfill  4.19 &\hfill 6.22  &\hfill 6.30  & \hfill\bf{6.32}  &\hfill 6.30  &\hfill 6.34   \\
\quad5    &          4.54$\pm$0.44 &\hfill 2.95 $\pm$ 0.15 &\hfill1.51 &\hfill  3.42 &\hfill 4.84  &\hfill 4.90  &\hfill \bf{4.91}  &\hfill 4.88  &\hfill 4.83   \\ 

\end{tabular}
\end{ruledtabular}

\label{tab:TCM_E0_parameter}
\end{table*}

\begin{equation}
    \sigma\left(PbPb \xrightarrow[]{1\gamma 1\gamma} Pb^*Pb^*\right) = \int P_{A_1}(b) P_{A_2}(b) d^2b, 
\label{eq:sigma_tot_double}    
\end{equation}
where
\begin{equation}
  P_{A_i}(b) = \int \sigma_{abs}\left( E_\gamma \right) N\left(E_\gamma,b\right) dE_{\gamma}.
\end{equation}

In our previous approach, Ref. \cite{Klusek-Gawenda:2013ema}, we implicitly  assumed
\begin{equation}
P(E_{exc};E_{\gamma}) \propto \delta \left( E_{exc} - E_{\gamma} \right)
\; ,         
\end{equation}
where $\delta$ is the Dirac delta function.
Above $P(E_{exc};E_{\gamma})$ can be interpreted as a probability 
of populating equilibrated compound nucleus with a given excitation 
energy $E_{exc}$ in a process initiated by the photon with energy 
$E_{\gamma}$. It must be constructed to fulfill the
following probabilistic relation:
\begin{equation}
\int_0^{E_{\gamma}} P(E_{exc}; E_{\gamma}) \;dE_{exc} = 1 \; .
\end{equation}

This relation can be generalized for a given number (k) of emitted neutrons. 
The agreement with the data for k = 1, 2 neutrons means that at small 
excitation energies $E_{exc}<30$~MeV, where the GDR mechanism dominates \cite{Klusek-Gawenda:2013ema}, 
the whole photon energy is absorbed by the nucleus. The deexcitation of the Pb nucleus is done with the GEMINI++ program \cite{gemini} which provides all information about $\gamma$-rays emission, various particle evaporation and fission in wide energy excitation range. 

The situation changes for larger excitation energies where not
the whole photon energy goes to the production of a equilibrated compound nucleus.

The energy could be dissipated in the pre-equilibrium emission of particles such as neutrons, protons or pions, the reactions of quasi-deuteron \cite{quasideuteron} or other processes.

In our TCM the $E_{exc}$ is typically smaller than the photon
energy $E_{\gamma}$. This was established already in
$\gamma + A$ collisions \cite{LEPRETRE1978}.
At small energies where giant dipole resonance is excited: 
$E_{exc} \approx E_{\gamma}$. Here the Compton scattering
is also possible in principle \cite{schumacher}. 
At larger energies only a part of energy is transferred to
the equilibrated system (see also \cite{reldis}).
This fact starts clearly at energy where reactions on quasi-deuteron
(or correlated $p n$ pair in another language)
become dominant. This means that already at these energies
we have preequilibrium processes. One (or even two) of 
the constituents of the $p n$ pair may escape before equilibration. At higher energies 
the situation is less obvious and the notion of preequilibrium 
is defined somewhat worse. The details depend on a particular 
dynamical model.
For energies $E_{\gamma} <$ 140 MeV one preequilibrium neutron may 
be a realistic case.
Our model was inspired by the presence of quasi-deuteron mechanism.
However, we will use it also for energies $E_{\gamma} >$ 140 MeV.


In order to better understand the situation, we consider
a simple model in which different excitation energies
$E_{exc} < E_{\gamma}$ can be populated. We started with the 
somewhat academic step-like function:
\begin{equation}
P(E_{exc}; E_{\gamma}) = \mathrm{const}(E_{exc}) = 1/E_{\gamma} \; 
\label{step-like}
\end{equation}
for $E_{exc} < E_{\gamma}$, i.e. uniform population
in excitation energy.

Another option is to take the simple sinus-like function:
\begin{equation}
P(E_{exc}; E_{\gamma}) = \frac{\pi}{2 E_{\gamma}} 
                    \sin \left( \pi E_{exc}/E_{\gamma} \right)
\label{sin-like}
\end{equation}
for $E_{exc} < E_{\gamma}$.

A comparison of the results with the two excitation functions 
will be instructive.

The following results for the interpretation of the excitation energy from 
the initial energy of the interacting photons are obtained with the
two-component model (TCM), where delta-like and step-like or 
sinus-like functions are combined.

In general, the bigger number of neutrons, the larger the 
fraction of energy carried by neutrons.

\begin{equation}
P(E_{exc}; E_{\gamma}) = 
                 c_1(E_{\gamma}) \delta \left( E_{exc}-E_{\gamma} \right)
               + c_2(E_{\gamma}) / E_{\gamma} \; .
               \label{two-component_model}
\end{equation}
The probabilistic interpretation requires:
\begin{equation}
c_1(E_{\gamma}) + c_2(E_{\gamma}) = 1 \; .
\end{equation}
In general, $c_1$ and $c_2$ may (should) depend on photon energy 
$E_{\gamma}$.
As a trial function for further analysis, we propose
\begin{eqnarray}
c_1(E_{\gamma}) &=& \mathrm{exp}\left( -E_{\gamma}/E_0 \right) \; ,
\label{eq:c1_E}
\\
c_2(E_{\gamma}) &=& 1 - \mathrm{exp} \left( -E_{\gamma}/E_0 \right) \; .
\label{eq:c2_E}
\end{eqnarray}
The parameter $E_0$ can be adjusted to the ALICE data \cite{ALICE502a}.
We suggest $E_0 \approx$ 50~MeV to start with.

The full TCM formula for the probability of emitting a specified number of neutrons for a given photon energy is calculated by:

\begin{equation}
\begin{split}
    P_k(E_{\gamma}) &= \sum_{E_{exc}}^{E_{\gamma}} ( 1-\mathrm{exp}(-E_{\gamma}/E_0) ) \frac{1}{E_{\gamma}} \frac{N_k(E_{exc})}{N_{ev}} \Delta E_{exc} \\
    &+ \mathrm{exp}(-E_{\gamma}/E_0) \frac{N_k(E_{\gamma})}{N_{ev}}.
\label{E0_param_eq}
\end{split}
\end{equation}

Here, the $N_k(E)$ is a number of events with k emitted neutrons for a given $E_{\gamma}$ or an excitation energy, and $N_{ev}$ is a total number of events for a given energy. Both numbers are obtained from GEMINI++ event generator. The $\Delta E_{exc}$ is a chosen interval of excitation energy in a discrete sum in Eq.~\ref{E0_param_eq}. This is purely technical parameter to simplify the calculations. The neutron emission probability fulfill the following condition:

\begin{equation}
    \sum_k P_k(E_\gamma) = 1.
\end{equation}

The formula (\ref{E0_param_eq}) contains a free parameter $E_0$, which was fitted to reproduce the dependence of the excitation energy of Pb on the photon energy beam.  Authors of \cite{guaraldo} displayed 
the dependence of the excitation energy on the photon energy 
using the cascade-evaporation model tuned to reproduce the data on fissilities of nuclei at intermediate energy.

In Table \ref{tab:TCM_E0_parameter} we show cross sections for the emission of various number of neutrons for different
values of $E_0$ parameter. The resulting cross sections for a given
number of neutron emissions (n = 1, 2, 3, 4, 5) only weekly depend on 
the value of the phenomenological parameter. Therefore in 
the following when referring to the TCM we will show results for 
$E_0$ = 50~MeV as representative. The bigger $E_0$ the higher cross section
for emission of 2, 3, 4, 5 neutrons. On the contrary, the emission
of single neutron decreases with increasing $E_0$.
The result for the sin-like function seems to be worse than the results for the step-like formula.
For comparison we display result for the pure GEMINI++ approach where $E_{\gamma}=E_{exc}$.
The pure GEMINI++ approach fails to describe the new ALICE data for neutron multiplicity
bigger than 2. Other more realistic functions can be proposed.


\subsection{Pre-equilibrium emission estimation}
\label{Sect.III}

Nuclei interacting with photons with energy less than (25~--~30)~MeV usually 
deexcite by neutron or $\gamma$-rays emission including the 
Giant Dipole Resonances. In this case all photon energy is transformed
into excitation energy of the nucleus. For more energetic photons (30~--~140)~MeV, the equilibration stage is long enough to open also the
possibility of the emission of the particles before thermalization of
the nucleus. In such a case, a part of the incident energy is lost by
neutron or/and proton emission, and the main deexcitation process starts from a
somewhat smaller nucleus than the initial colliding one.

In the moderate energy regime, below the particle production threshold,
there exist some theoretical estimation how much energy can be used for
the pre-equilibrium stage. 

The photon energy regime above 140~MeV is connected with the highly energetic 
particles production such as pions, muons and it is quite well described, e.g. 
by the GiBUU model \cite{GiBUU}.

In the review of Pshenichnov et al. \cite{reldis}, the results of the RELDIS  model are presented. The author explains the difference between the excitation and incident energies by the pre-equilibrium emission of the particles.

This assumption encourage our study on the estimation of the amount of the particles lost by the nucleus during thermalization process. Thus in place of TCM one can try to use another realistic predictions of 
the particles emitted before the energy equilibration of nucleons 
inside the excited nucleus occurs. One of the possibility is 
to employ the well-known collisions models relevant for
particle+nucleus collisions such as Heavy Ion Phase-Space Exploration (HIPSE) \cite{hipse1}.

The HIPSE model describes the heavy-ion collision in several stages: from the contact point, by reaggregation of the particles coming from overlapping nuclei till deexcitation of the thermalized prefragment. This last stage is done within the state-of-art statistical approach GEMINI++. 
In general, it was used with success to describe reaction with collision energies  15~--~100~MeV/nucleon. For the sake of this study, the reaction n+$^{207}$Pb is chosen, as the neutron interactions should be the best possible approximation of the photon interaction one can do.\footnote{The HIPSE does not allow for $\gamma$+A collisions.}

\begin{figure}[h]
    \centering
        \includegraphics[width=9.3cm]{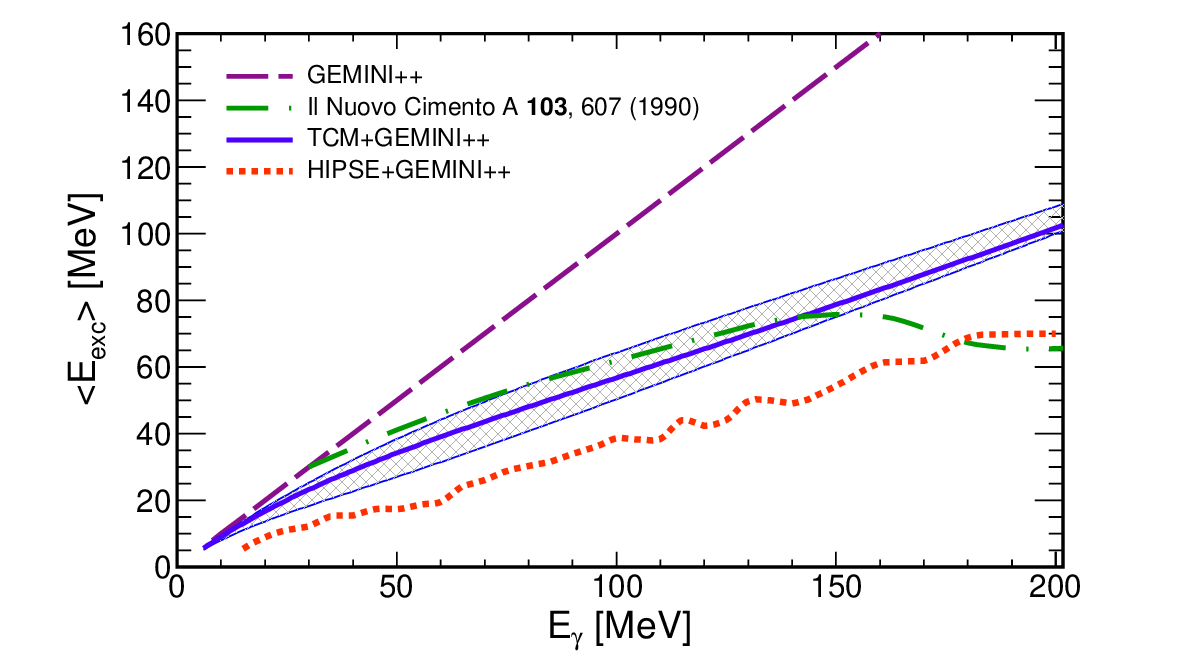}\vspace{-0.cm}
    \caption{ Dependence of excitation energy of $^{208}$Pb ion as a function of photon energy. The Monte Carlo CASCADE simulation \cite{guaraldo} (green), the TCM (blue), GEMINI++ (purple) and HIPSE+GEMINI++ (red). The shadowed area shows results of TCM with $E_0\in (20,80)$~MeV.}
    \label{fig:energy}
\end{figure}

Fig.~\ref{fig:energy} allows to estimate how much energy is dissipated into internal degrees of freedom in various models. Previous excitation energy estimations were done with the help of intranuclear cascade to describe experimental photofission cross sections, for Bi and Au nuclei \cite{guaraldo} and also average multiplicities of neutrons, protons and pions for Pb and U \cite{barashenkov} in photon energy range 0.05-1.0~GeV. 
The HIPSE calculation gives a similar trend. The average excitation energy in TCM with step-like function can be approximated as:
\begin{equation}
    <E_{exc}> = c_1(E_{\gamma}) \cdot E_{\gamma} + c_2(E_\gamma) \cdot \frac{E_\gamma}{2}.
\end{equation}
\noindent The photon of 100~MeV energy excites the Pb nucleus mostly to 40~MeV and the rest is lost on emission of pre-equilibrium particles and internal rearranging of the nucleons inside the hit nucleus. The shadowed area shows the results of fitting $E_0$ parameter (Eq.~(\ref{eq:c1_E}, \ref{eq:c2_E})) and in average the $E_0=50$~MeV gives the best reproduction of data from previous calculation \cite{guaraldo}. Thus, this value is taken to further investigation.

\begin{figure}
    \includegraphics[width=8.5cm]{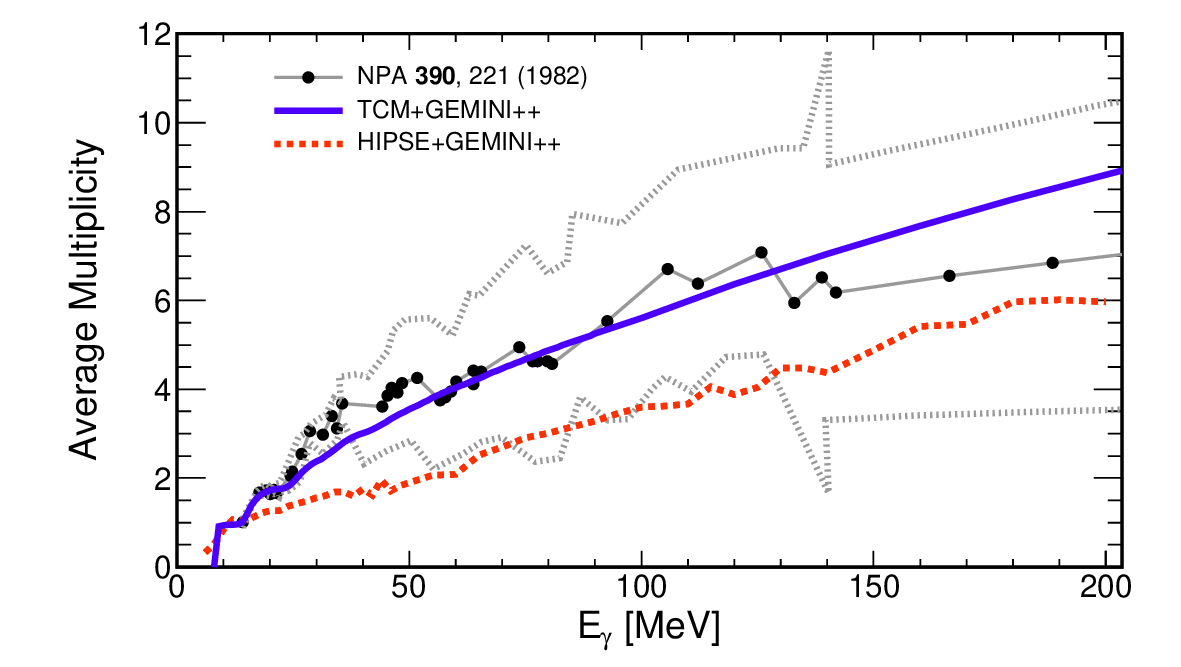}
    \caption{The mean neutron multiplicity: experimental \cite{LEPRETRE1982} (black) and estimated with TCM (blue) and HIPSE model (red). The grey dotted lines show the experimental uncertainty.}
    \label{fig_km2}
\end{figure}

The difference between neutron and photon-induced reactions are displayed in Fig.~\ref{fig:energy}. The Guaraldo et al. 
\cite{guaraldo}, the TCM and the HIPSE models estimations of excitation energy show that the energy dissipation ($E_{\gamma}-E_{exc}$) in the case of the neutron-induced reaction can increase up to 80~MeV for $E_{\gamma}$ = 150~MeV in comparison to photon-induced one, where it is around 60~MeV. This is also well visible in the case of the mean neutron multiplicity presented in Fig.~\ref{fig_km2}.
The experimental points were extracted from a measurement of photon-induced neutron emission \cite{LEPRETRE1978}.
The average number of neutrons is still within the experimental uncertainty but missing around 2 neutrons in the case of HIPSE+GEMINI++. Thus these event generator estimations give only the lower limit of the amount of particles lost during the process of equilibration of an excitation energy between nucleons inside the colliding nucleus.

The performance of the TCM average multiplicity is much better, which is obvious as the function has been fitted to the previously checked dependence of excitation energies.  
 Despite this small inconsistencies, the HIPSE allows to verify the
 statement that for reaction with photon energies larger than 30~MeV, the pre-equilibrium emission is very important. The maximal number of neutrons emitted during one event changes from 0.9 for $E_{\gamma}$ = 30~MeV up to 11 for $E_{\gamma}$ = 140~MeV, thus this is the minimal amount of particles which the initial nucleus loses during the energy equilibration stage.
 
 In general, not only neutrons are evaporated.
\begin{figure}
    \includegraphics[width=8.9cm]{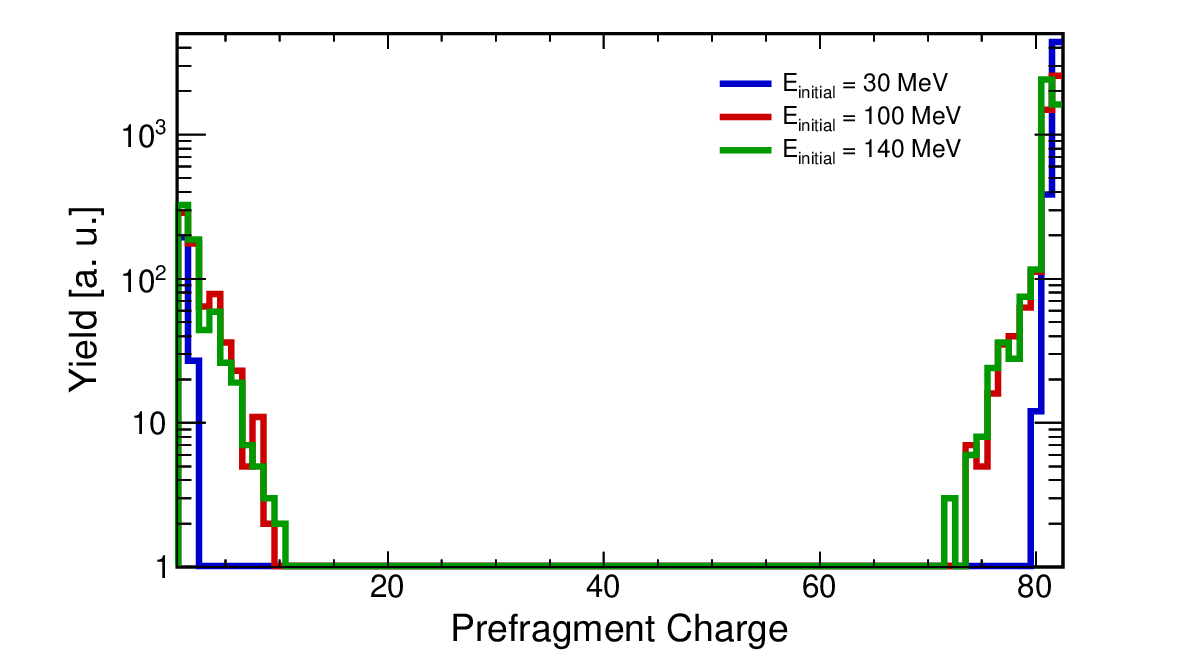}
    \caption{The prefragment charge distribution calculated with the HIPSE model for neutron incident energy 30, 100 and 140~MeV.}
    \label{fig_km3}
\end{figure}
Fig.~\ref{fig_km3} shows the distribution of the charge of the
prefragments left after the equilibration stage. 
The initial $^{208}$Pb emits up to 10 nucleons for $E_{\gamma} >$ 
100~MeV. For low-energy reaction, there are mainly neutron and proton 
evaporation channels opened, but for 140~MeV incident energy even,
isotopes of carbon and oxygen could reduce the charge of the initial lead nucleus.
 
The present studies may also be important for a better understanding 
of the electromagnetic interaction between the spectator and 
particles produced in the participant zone \cite{Mazurek1, Mazurek2}.

It is quite surprising that even for the ultraperipheral collision 
the charge of the nucleus can differ by as many as 10 units (Fig.~\ref{fig_km3}), thus in the case of lead nuclei, there is a full ensemble of species produced down to Ytterbium. Of course, such cases have a low probability but still it has to be taken into consideration.

 \begin{figure}
    \includegraphics[width=8.5cm]{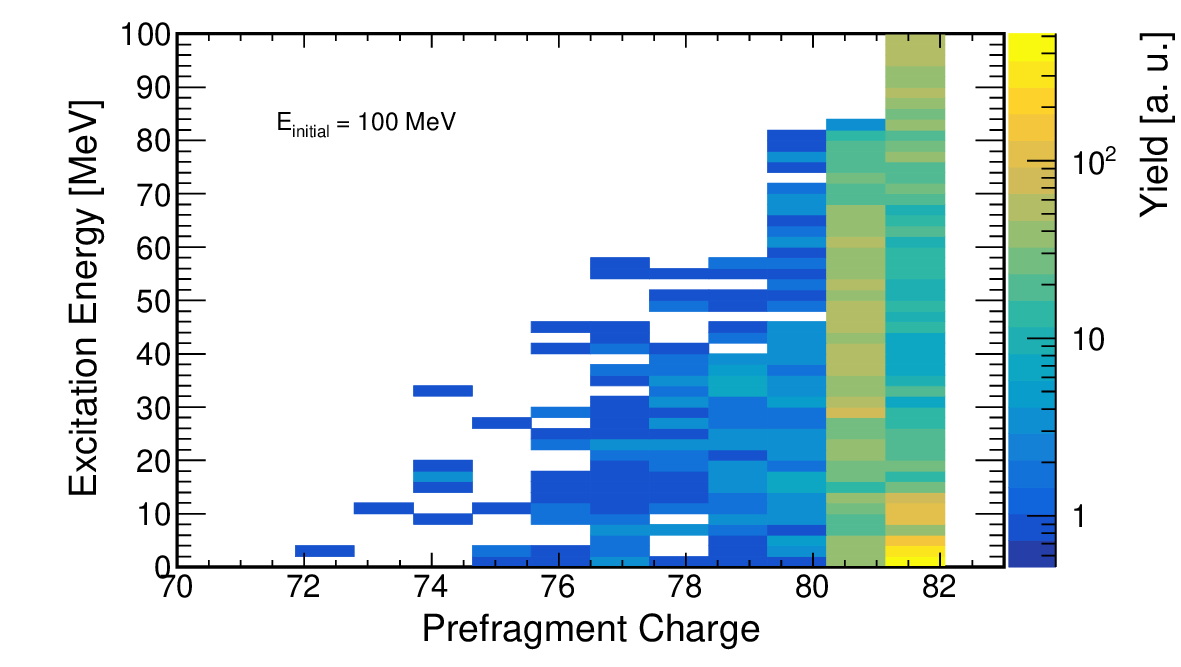}
    \caption{The distribution of the excitation energy of the prefragment charges calculated with the HIPSE model for neutron incident energy 100~MeV.}
    \label{fig_km4}
\end{figure}

Moreover, the predicted excitation energy for a given incident energy (i.e. E$_{initial}$~=~100~MeV in Fig.~\ref{fig_km4}) is distributed in full range and depends on the mass and charge of the prefragments. The emitted particles could carry away kinetic energy in addition to the energy needed to separate them from the excited nucleus. 

\subsection{Compound nucleus deexcitation}
\label{Sect.IV}
The deexcitation of the hot nucleus at moderate excitation energy can be described by many statistical and dynamical models. In the present studies the state-of-art statistical model GEMINI++ \cite{gemini} was employed as it gives very reasonable results in the investigated energy regime. Moreover, in the past, it was used to discuss the Giant Dipole Resonance emission from the photon-induced Pb collision below 30~MeV \cite{Klusek-Gawenda:2013ema}.
As the GEMINI++ doesn't depend on the way of producing hot nuclei, but only on the initial excitation energy, spin, mass and charge are important, it can be also used as an afterburner.


The evaporation process in the GEMINI++ statistical code~\cite{gemini,Cie15} is described by the Hauser-Feshbach formalism~\cite{hauser}, in which the decay width (in MeV) for the evaporation of an $i$-th particle from the compound nucleus, with an excitation energy of $E^*$ and spin $s_{CN}$, is given by the expression
\begin{eqnarray}
     &&\Gamma_i = \frac{1}{2\pi \rho (E^*, s_{CN})} \times
                                                               \nonumber \\
     &&\int d \epsilon
      \sum_{s_d=0}^{\infty}
      \sum_{J=|s_{CN}-s_d|}^{s_{CN}+s_d}
      \sum_{\ell=|J-s_i|}^{J+s_i}
            T_\ell (\epsilon) \rho(E^*-B_i - \epsilon, s_d).   \qquad
                                                               \label{eqn.01}
\end{eqnarray}
Above: $s_d$ is the spin of the daughter nucleus, while $s_i$, $J$, and $\ell$ are, respectively, the spin, total angular momentum, and orbital angular momentum of the evaporated particle, $\epsilon$ and $B_i$ are the kinetic and separation energies, $T_\ell$ are the transmission coefficients, and $\rho$ and $\rho_{CN}$ are the level densities of the daughter and compound nuclei. They have been calculated using the expression taken from~\cite{gemini}:
\begin{eqnarray}
      \rho(U,s)
      &=&
      \frac{(2s + 1)}{24\sqrt{2} (1+ U^{5/4}\sigma^3) \sqrt[4]{a(U,s)}}
                                                               \nonumber \\
      &\times&  \mathrm{exp}(2\sqrt{a(U,s)U}),
                                                               \label{eqn.02}
\end{eqnarray}
where $\sigma= \sqrt{\mathcal{J}T}$, with $\mathcal{J}$ being a moment of inertia of a rigid body with the same density as the nucleus. In this context,  $T$ is the nuclear temperature defined as:
\begin{equation}
     \frac{1}{T}
     =
     \frac{d S}{d U},
     \;\;\leftrightarrow\;\;
     S = 2 \sqrt{a(U,s) U},
                                                               \label{eqn.03}
\end{equation}
where $S$ represents  the nuclear entropy, and where  thermal excitation energy $U$ is calculated with taking into account both, pairing $\delta P$ and the shell corrections $\delta W$:
\begin{equation}
     U = E^* - E_{\rm rot}(s)  + \delta P +\delta W.
                                                               \label{eqn.04}
\end{equation}
Here, $E_{\rm rot}(s)$ stands for  the rotational energy of the nucleus.

The level density parameter $a(U,s)$ was parametrized after Ref.\,\cite{gemini} as
\begin{equation}
      a(U,s)
      =
      \tilde{a}(U)
      \left(
            1 - h(U/\eta + s/s_\eta)\frac{\delta W}{U}
      \right),
                                                               \label{eqn.05}
\end{equation}
where $\delta W$  is the shell correction to the liquid-drop mass and
$\tilde{a}$ is the smoothed level-density parameter (see below). The separation energies $B_{i}$, nuclear masses, shell and pairing corrections were taken from Ref.~\cite{moller}.  The function specifying the rate of hindrance is $h(x) = \tanh{x}$, with damping  parameters  $\eta =$ 18.52~ and $s_\eta = $ 50$\hbar$~\cite{gemini}.

Level density parameter $\tilde{a}(U)$ depends on the excitation energy:
\begin{equation}
     \tilde{a}(U)
     =
     \frac{A}{k_{\infty} - (k_\infty -k_0)
     \mathrm{exp}\left(-\frac{\kappa}{k_\infty - k_0}\frac{U}{A} \right)},
                                                               \label{eqn.06}
\end{equation}
with parameters: $k_0=7.3$~MeV, $k_{\infty}=12$~MeV and   $\kappa = 0.00517 \mathrm{exp}(0.0345 A) $~\cite{gemini}.

Within the Monte Carlo \textsc{GEMINI++} calculations we assume that the excited nucleus was formed with angular momentum equal to 0~$\hbar$~(which is a good approximation for photon-induced reaction) and its excitation energy which was probed with 2~MeV step for energies below 100~MeV and with 20~MeV step for above 100~MeV value (with generated $10^6$ events per every excitation energy).

\subsection{Model results}


{The present investigation compares four scenarios: \\
\linespread{1.5}
(1) The pure GEMINI++ assuming $E_{exc}=E_{\gamma}$;\\
\linespread{1.5}
(2) The GEMINI++ results weighted by the Dirac delta + step-like functions (TCM) which nicely reproduces previous theoretical and experimental results of \cite{guaraldo, LEPRETRE1981};\\
\linespread{1.5}
(3) The hybrid model, where the collision part is described by n+Pb reactions within HIPSE event generator and next cooled down with GEMINI++; \\
\linespread{1.5}
(4) EMPIRE - Nuclear Reaction Model Code System for Data Evaluation \cite{empire} - set of nuclear models which permits to calculate various nuclear reaction in a broad range of beam energies. \\
\linespread{1.5}}

\begin{figure}[h]
    \includegraphics[width=9cm]{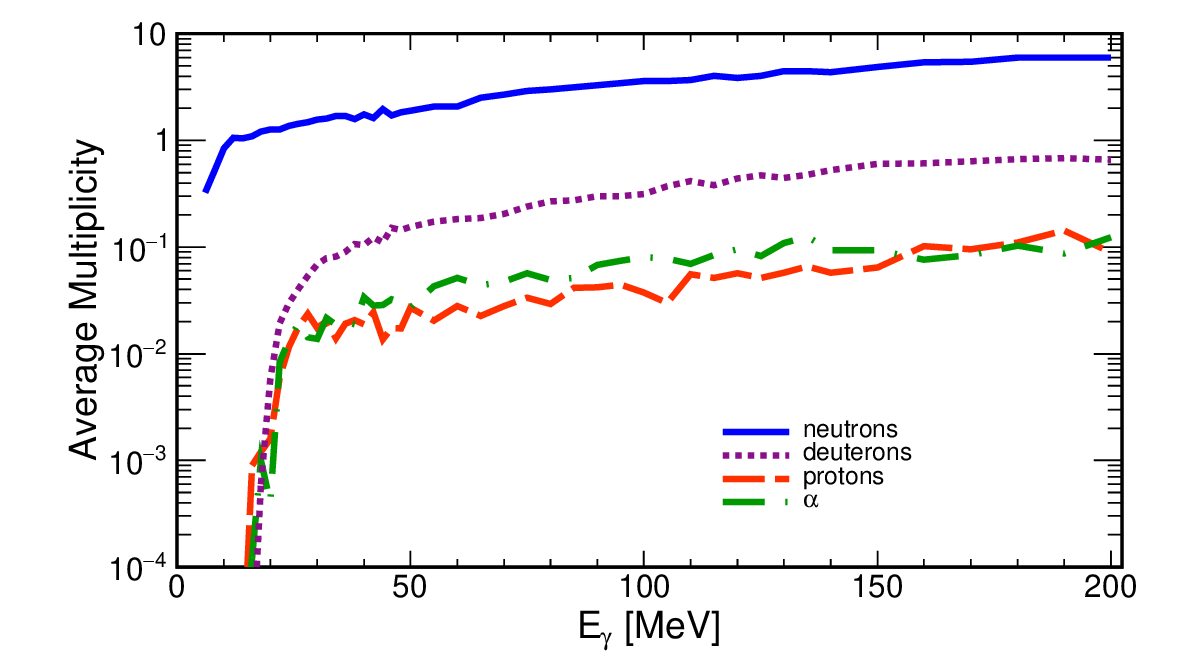}
    \caption{The average multiplicity of the neutron, proton, deuteron and $\alpha$ particle emitted during the full process (including pre-equilibrium stage) calculated with HIPSE + GEMINI++ hybrid model.}
    \label{fig_km5}
\end{figure}

Fig.~\ref{fig_km5} gives the idea about the mean multiplicity of neutrons, protons and alpha particles, emitted during the whole process. It is obvious that neutrons are the most favorable channel of deexcitation but still some numbers of protons and alphas but also deuterons, tritons and other light charged particles could be detected. 

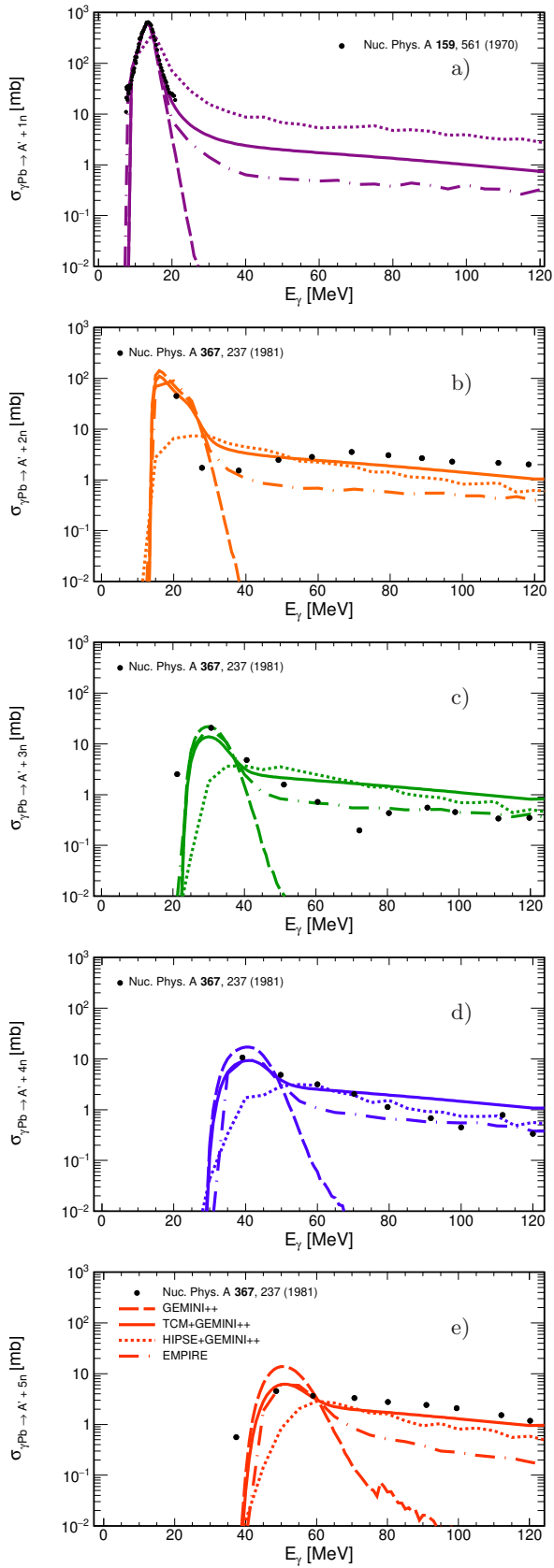
\begin{figure}[H]
 \centering
 \vspace{1cm}
 \setlength{\unitlength}{0.1\textwidth}
\begin{picture}(5,12)
\put(0,10){\includegraphics[width=8cm]{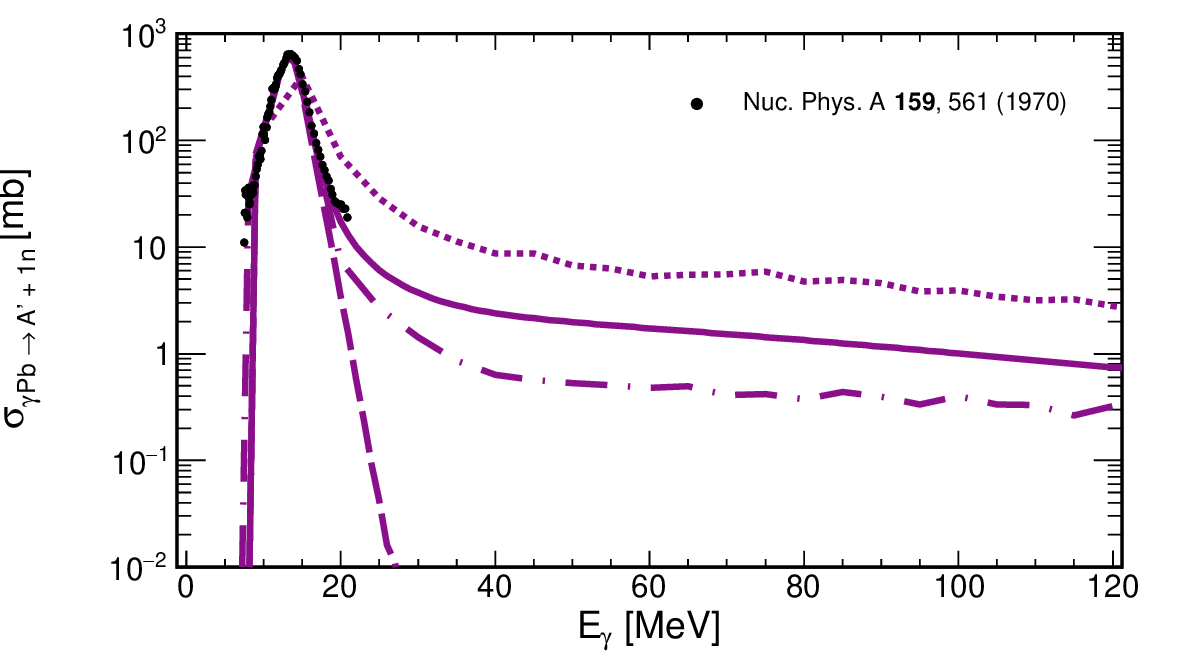}}
\put(3.5,11.9){a)} 
    \put(0,7.5){\includegraphics[width=8cm]{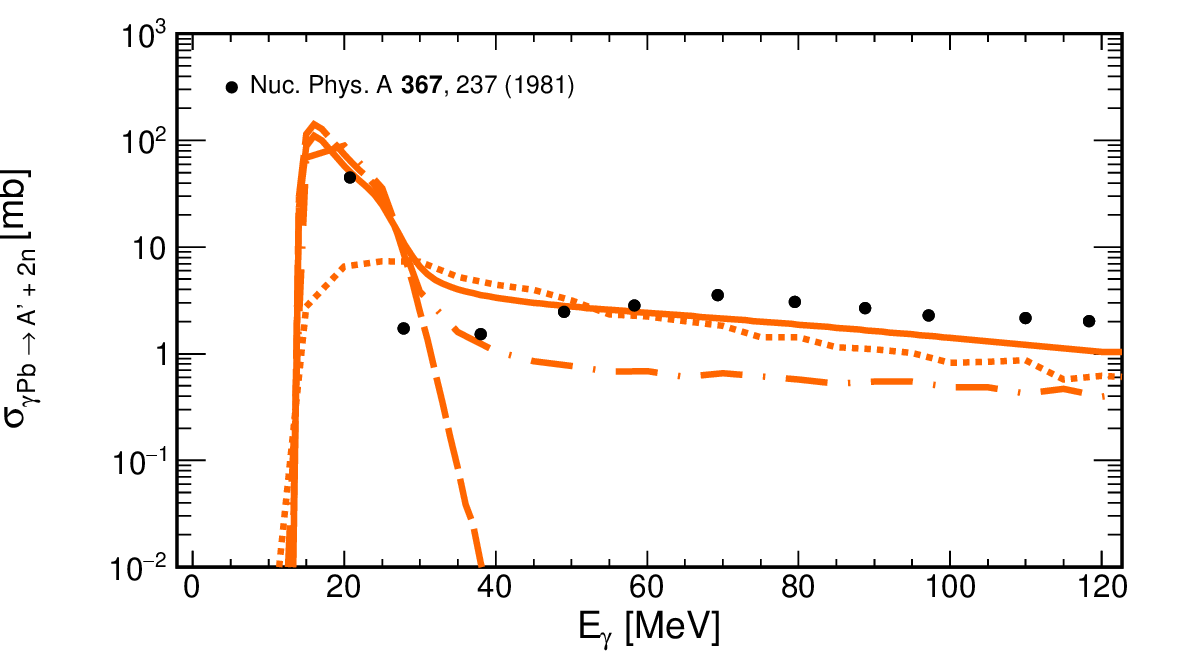}}
\put(3.5,9.4){b)}  
    \put(0,5){\includegraphics[width=8cm]{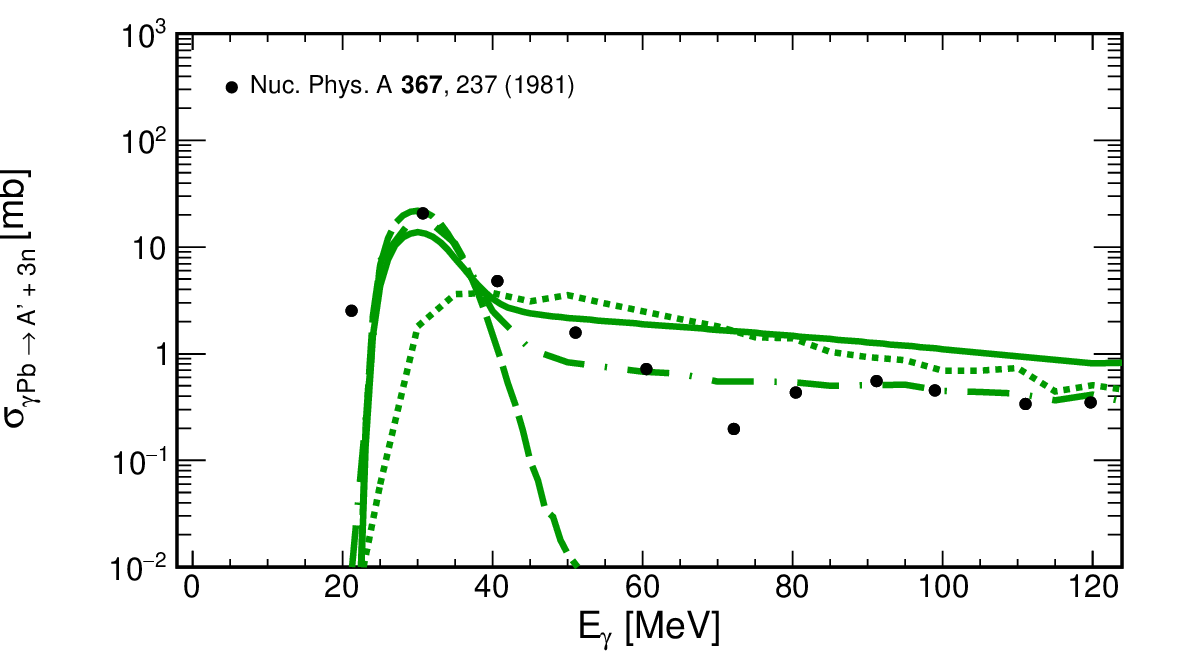}}
\put(3.5,6.9){c)}
     \put(0,2.5){\includegraphics[width=8cm]{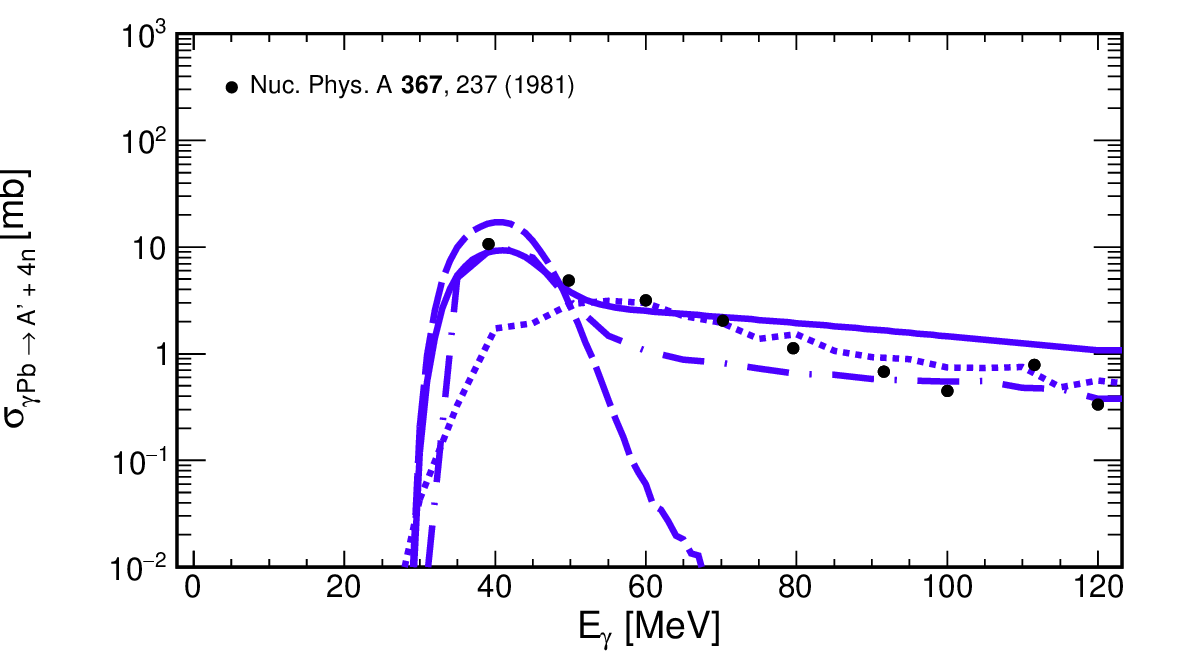}}
\put(3.5,4.4){d)}
    \put(0,0){\includegraphics[width=8cm]{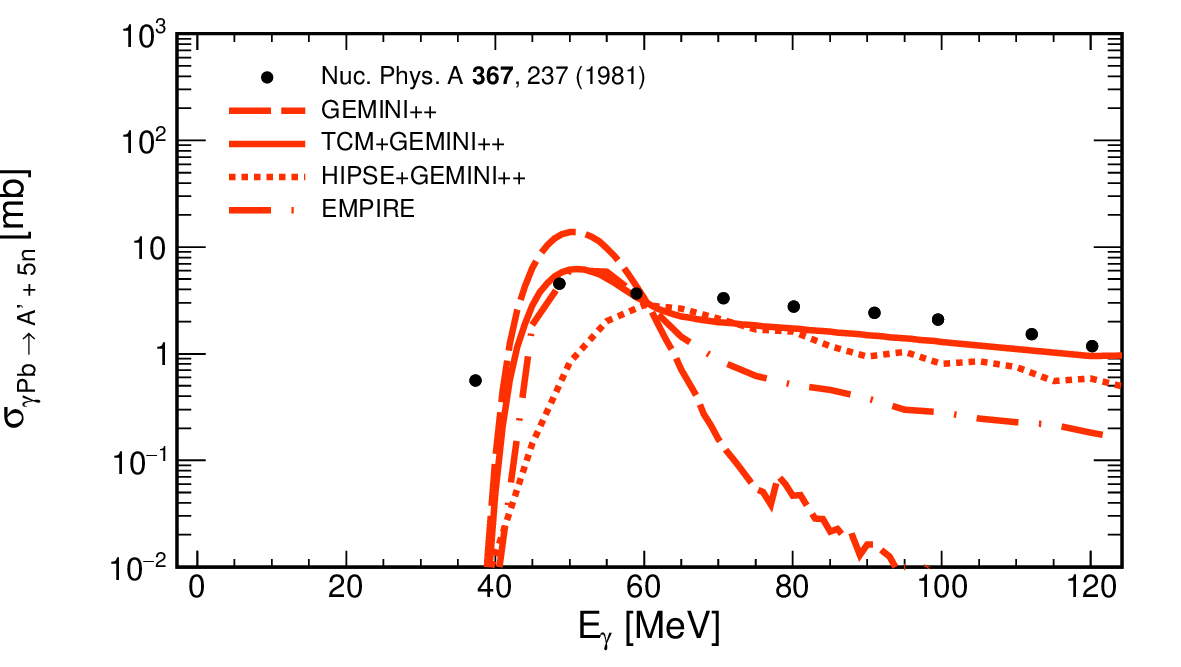}}
    \put(3.5,1.9){e)} \end{picture}
    \caption{The cross section for the $\gamma + ^{208}Pb \to kn + X$ reaction, where are displayed neutron multiplicities: k=1 (a), k=2 (b), k=3 (c), k=4 (d) and k=5 (e). We compare results of experimental data from \cite{LEPRETRE1981} (dots) 
    with model calculations: pure GEMINI++ (dashed line), TCM+GEMINI++ (solid line), HIPSE+GEMINI++ (dotted line) and EMPIRE (dash-dotted line). The color coding here will be used in the rest of the paper.
    }
    \label{fig:gammaA_knX_exp}
\end{figure}

The measurements of a given number of neutrons coming from 
deexcitation of photon-induced excitation of Pb in the $E_\gamma$ energy range 
(30~--~140)~MeV are very seldom. A set of articles on this subject was presented 
by Lepretre et al. in the early nineties \cite{LEPRETRE1981,LEPRETRE1982}. In their experiments the total nuclear absorption cross section for Sn, Ce, Ta, Pb and U was obtained using a monochromatic photon beam.
The experimental data for given neutron numbers shown in Fig.~\ref{fig:gammaA_knX_exp} are extracted from a Fig.~2 in Ref.~\cite{LEPRETRE1981}, where so-called cumulative results were shown. 

Assuming that the full photon energy is transformed into excitation
of $^{208}$Pb the full line shows pure GEMINI++ neutron energy spectra in
Fig.~\ref{fig:gammaA_knX_exp} which reproduces nicely the low energy bump
in cross section for each neutron number distribution but is missing the
high energy tails. The TCM + GEMINI++ (dashed line) and HIPSE + GEMINI++
(dotted line) reproduce high $E_\gamma$ the tails but HIPSE underestimates the low 
energy neutron production. 

In the first approach we assume that the whole photon energy is absorbed by the second
(absorbing) nucleus i.e. $E_{exc} = E_{\gamma}$ and assume equilibrium.
Then, the emission of neutrons is given by a cascade
of the Hauser-Feshbach emissions as implemented in the well-known
code GEMINI++ \cite{gemini}.

The best data reproduction was obtained by EMPIRE \cite{empire} calculations. This is due to the fact that EMPIRE combines many different nuclear models suited to correctly describe the specific phenomena and fit some selected experimental data. 

The photon-induced reaction calculation includes the initial photo-nuclear excitation process, subsequent decay of the excited nucleus by particle and $\gamma$-ray emission. The Giant Dipole Resonance and photo-absorption on a neutron-proton pair (a quasi-deuteron) are also taken into account. The Hauser-Feshbach theory is used for describing the sequential decay of hot nuclei similar to the GEMINI++ approach.

\section{Neutron emission at higher photon energies}

In our calculations we use $\gamma$A photoabsorption cross section as parametrized in \cite{Klusek-Gawenda:2013ema}. As a consequence, for $\sqrt{s_{NN}}$ = 5.02 TeV we get single-photon dissociation cross section $\sigma_{ED}$ = 210.4 b. This is similar to the estimation in \cite{cs5}. It was already recognized in Ref.~\cite{cs5} that the integration must be done up to very high photon energies.

If $E_{\gamma} >$ 140~MeV the situation changes again
as nonnucleonic degrees of freedom (nucleon resonances, partons) must 
enter the game.
Here the intra-nuclear hadronic cascade model(s)
(see e.g. \cite{Bertini1963,Bertini1969,FLUKA1, FLUKA2}) seem to be more 
appropriate than nuclear neutron evaporation of the Hauser-Feshbach 
type discussed in the previous section.\footnote{Even at low energies the standard neutron evaporation
is modified by the reactions on quasi-deuteron \cite{Levinger}.}
Most of the INC codes are written rather for hadron-nucleus
interactions and not for photon-nucleus interaction.
The photon-nucleus interactions are, however, slightly different
and, in principle, require a dedicated implementation.


In general, the Eq.~(\ref{eq:sigma_tot_neutron_sum}) can be developed to obtain the cross section for production of neutrons in UPC 
in one-photon exchange approximation as follow:
\begin{eqnarray}
&&\sigma_{A A \to A A' + k n}  = \nonumber\\
&&\int d^2 b_1 d E_{\gamma} 
\sigma_{\gamma + A \to k n + A'}(E_{\gamma} ) W_1(b) \frac{d N}{d^2 b_1 d E_{\gamma} } \; .
\label{one-photon_approximation}
\end{eqnarray}
Above $\sigma_{\gamma + A \to k n + X}$ is the cross section for
inclusive emission of $k$ neutrons, i.e. is integrated over
directions of each of the emitted $k$ neutrons. 
$W_i(b) \approx \frac{exp(-m(b))}{i!}$ is the weight factor obtained for $m(b)$ - average number of absorbed photons \cite{reldis}.
In \cite{NooN} those cross sections are taken directly from
experimental data (see e.g.~\cite{Mainz}) and extrapolated
to high energies assuming constant (photon energy independent) 
cross section. The experimental data are limited to  
$E_{\gamma} <$~140~MeV.
Is this extrapolation to higher $E_{\gamma}$ supported by 
the underlying dynamics? What is the underlying dynamics?

The multiplicity dependent cross section

\begin{equation}
\sum_{k=0}^{N_{max}} \sigma_{\gamma+ A \to k n + A'}(E_{\gamma} )
\approx \sigma^{tot}_{\gamma + A}(E_{\gamma} ). 
\end{equation}
Somewhat formally one can write each term in $k$,
corresponding to the emissions of $k$ neutrons, as:
\begin{equation} 
 \sigma_{\gamma+ A \to k n + A'}(E_{\gamma} )  =
 \sigma^{tot}_{\gamma+ A}(E_{\gamma} ) P_k(E_{\gamma} ) .
\end{equation}

Alternatively one can therefore parametrize $P_k(E_{\gamma} )$ function
to include tails at large $E_{\gamma}$.
We do not have, however, reliable predictions as a high-energy
$\gamma+ ^{208}Pb \to k n + A'$ was not studied experimentally.
As a simplest ``solution'' we assume $P_k(E_{\gamma} )$ function to be 
a constant in $E_{\gamma} $. A better approach would be an explicit
dynamical model. One possibility would be to use
the GiBUU transport approach \cite{GiBUU}. The calculation
in \cite{LS2020} shows that for large energies of quasi-real photons
both small ($M_n <$ 5) and large ($M_n >$ 10) neutron multiplicities 
are possible (see Fig.~4 there). Even at large energies the cross 
section for the emission of a small number of neutrons stays large.

\begin{figure}[bt]
\includegraphics[width=8cm]{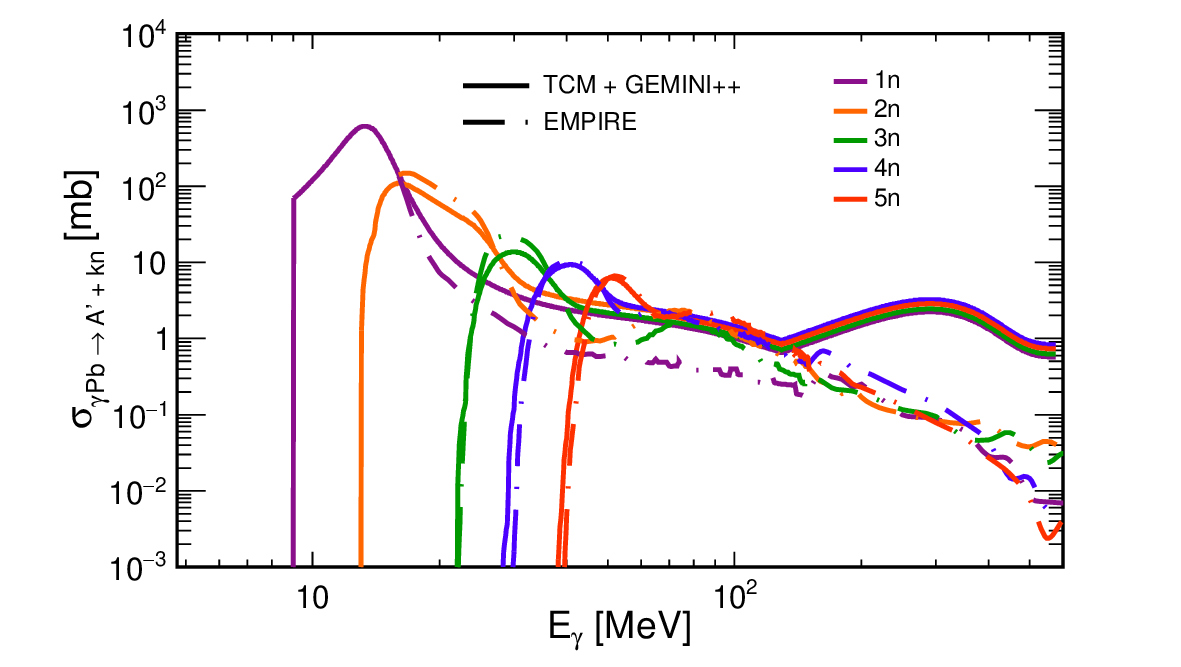}
\caption{Cross section for $\gamma+ ^{208}Pb \to kn + A'$ as a function of photon energy.
The solid line is for our TCM ($E_0 = 50$~MeV) with constant probabilities
$P_k$ above $E_\gamma$ = 200~MeV. One may clearly see the presence of the $\Delta$ resonance
at $E_{\gamma} $~$\approx$~200~MeV which is the desired feature of our approach.
For comparison (dashed lines) we show similar results from the EMPIRE code.}
\label{fig:energy_dependence_for_fixed_multiplicity_models}
\end{figure}

In Fig.~\ref{fig:energy_dependence_for_fixed_multiplicity_models}
we show the cross section for $\gamma+ ^{208}Pb\rightarrow kn +X$ as a function of the photon energy, 
from the threshold (a few~MeV) to 500~MeV,
for different multiplicity of neutrons. The results of 
the EMPIRE code \cite{empire} (short dashed lines) are compared with our TCM extrapolation 
with constant $P_k(E_{\gamma} )$ (solid lines).
Above $E_{\gamma} >$ 140~MeV the two approaches diverge.
Our TCM with constant $P_k$ shows a maximum at 
$E_{\gamma} \sim$ (300~--~400)~MeV which is due to excitation of nucleon 
resonances, mostly $\Delta$ isobars.

The higher energies, 
especially in the region of partonic excitations, require further 
theoretical studies in future.
There was some discussion on high-energy scattering of photons on 
nuclei within dual parton model in \cite{RER1996,RER2001}. 
According to our knowledge, no neutron emission was studied in this context.

What are photon energies responsible for emission of 1-5 neutrons is a crucial dynamical problem which we approach here in a very phenomenological way.

\begin{figure}[bt]
 \centering
 \setlength{\unitlength}{0.1\textwidth}
\begin{picture}(5,7.5)
\put(0,5){\includegraphics[width=8cm]{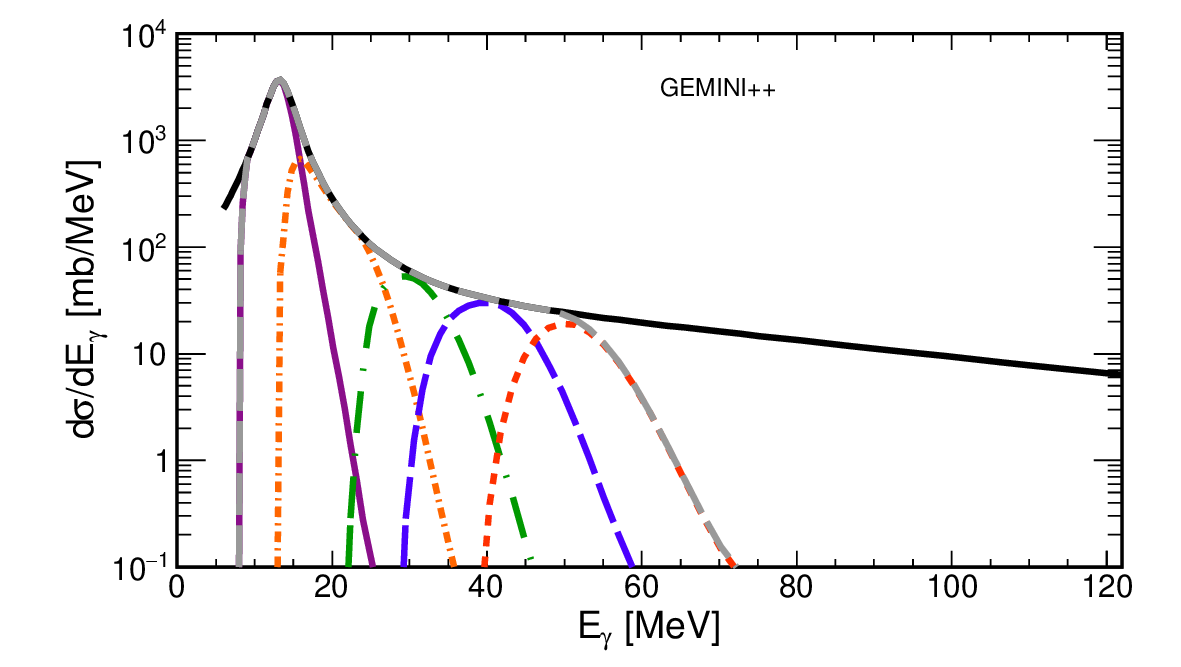}}
\put(2.,7.){a)} 
    \put(0,2.5){\includegraphics[width=8cm]{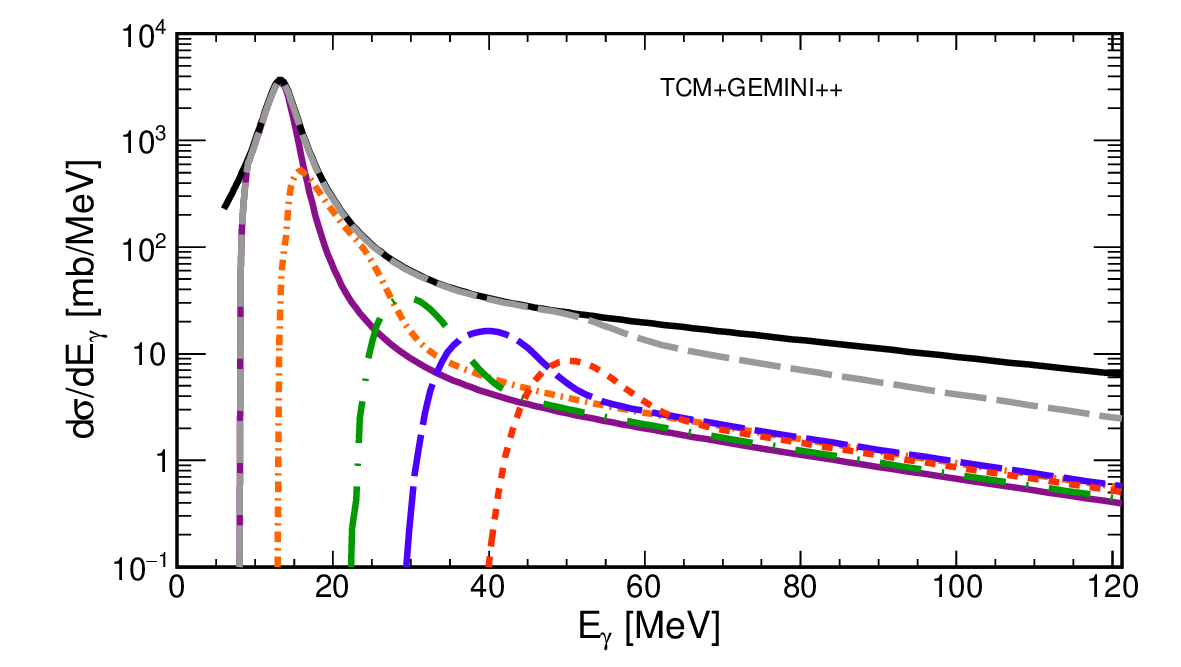}}
\put(2.,4.5){b)}  
    \put(0,0){\includegraphics[width=8cm]{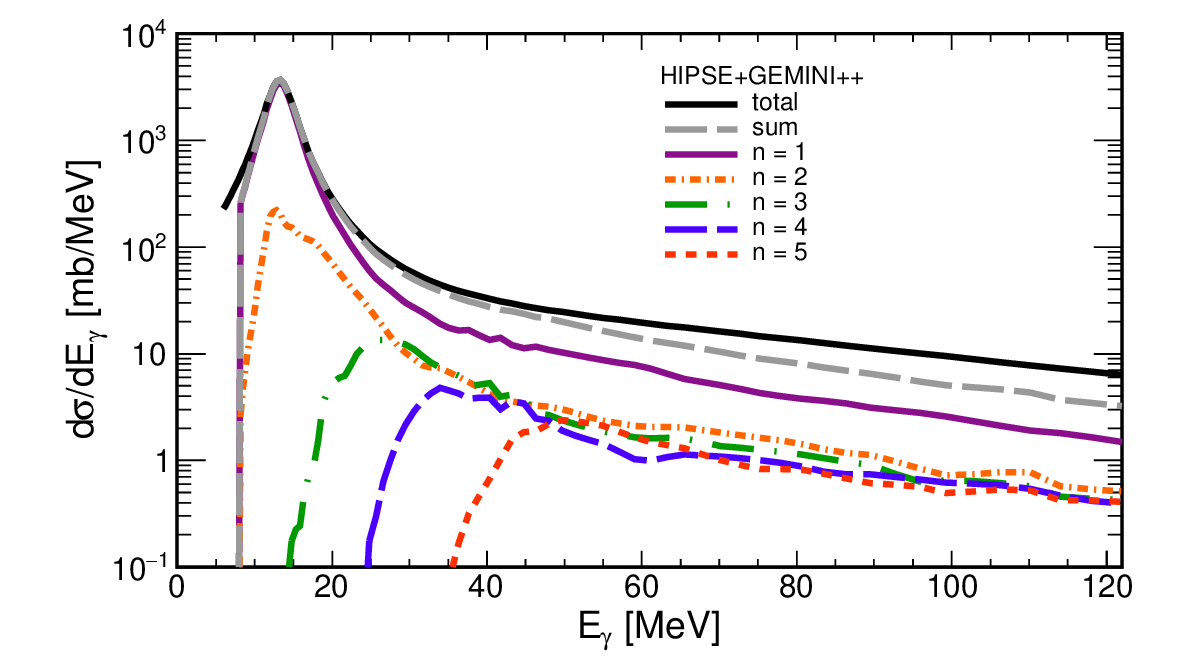}}
\put(2.,2.){c)}
 \end{picture}
  \caption{$E_\gamma$ distribution for a fixed number of neutron
     emissions for
     UPC with energy $\sqrt{s_{NN}}$= 5.02~TeV calculated via (a) GEMINI++, (b) TCM+GEMINI++ and (c) HIPSE+GEMINI++. The solid black line is
     obtained from Eq.~(\ref{eq:sigma_tot_neutron_sum}) while the sum of
     up to 5 neutron contributions by the dashed line.
     }
    \label{fig:UPC_hipse}
\end{figure}
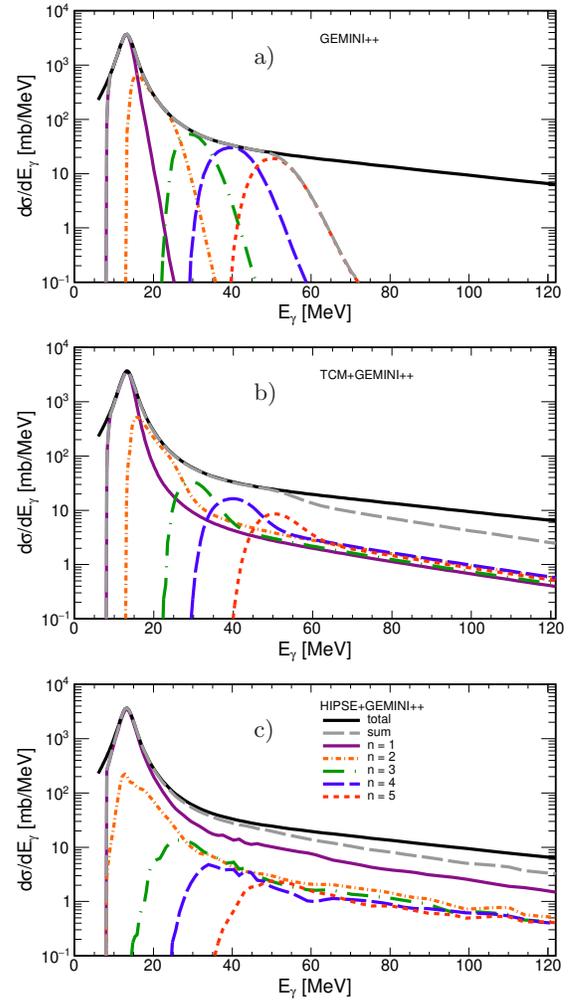

The outcome of Eq.~(\ref{one-photon_approximation}) is displayed in
Fig.~\ref{fig:UPC_hipse} where the distribution of 1, 2, 3, 4 and 5 neutron emissions as a function of photon energy is shown. The black line marks the total cross section for neutron emission. 
Fig.~\ref{fig:UPC_hipse}(a) is for pure Hauser-Feshbach approach as encoded in GEMINI++, Fig.~\ref{fig:UPC_hipse}(b) relates to the TCM approach and Fig.~\ref{fig:UPC_hipse}(c) to HIPSE+GEMINI++ model.

\begin{figure}[bt]
 \centering
 \setlength{\unitlength}{0.1\textwidth}
\begin{picture}(5,5)
    \put(0,2.5){\includegraphics[width=8.5cm]{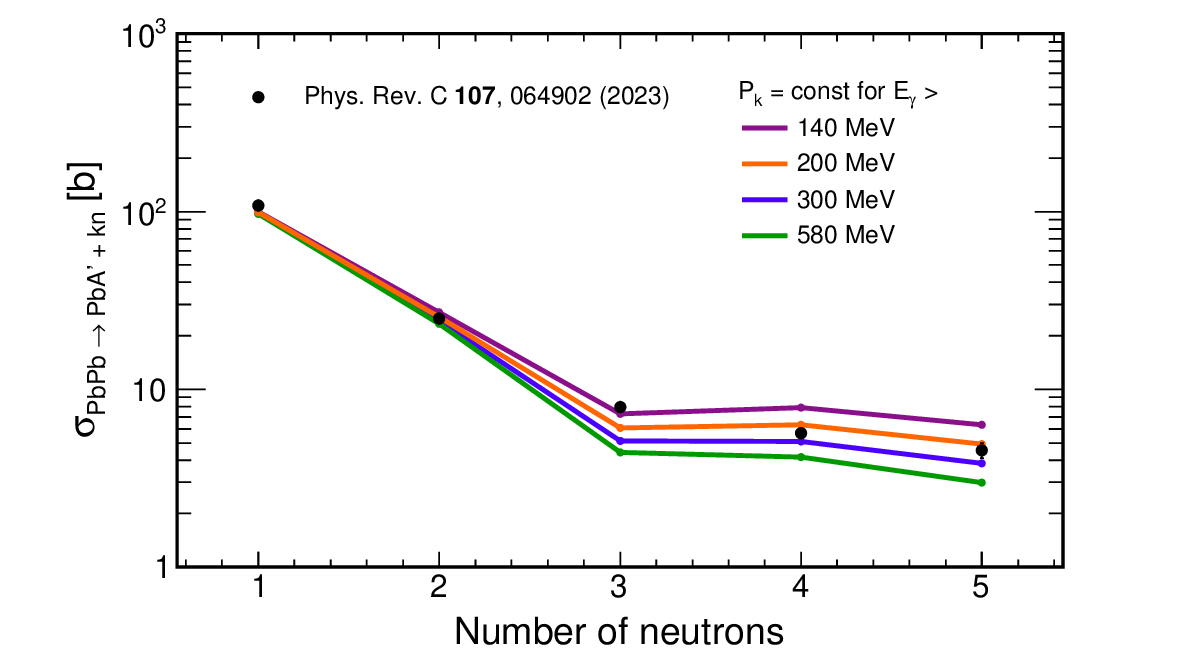}}
\put(1.,4.5){a)}  
    \put(0,0){\includegraphics[width=8.5cm]{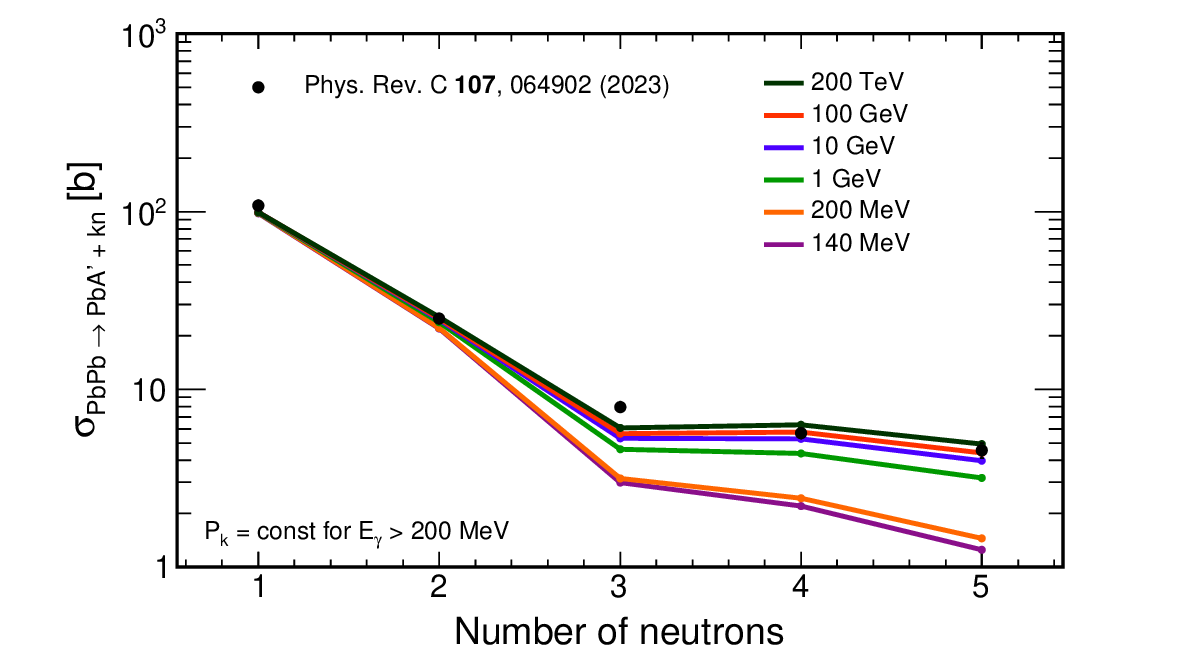}}
\put(1.,2.){b)}
 \end{picture}
\caption{ Cross section for emission of a number of neutrons for
different upper limits on $E_{\gamma} $. In panel a) the constant value
of $P_k$ starts for changing value given in the panel. The upper
limit is very high. In panel b) the $P_k$ constant value starts for
$E_\gamma$ = 200~MeV. The upper limit is modified to the values
shown in the panel.}
\label{fig:sigma_tot_k}
\end{figure}

In Fig.~\ref{fig:sigma_tot_k} we present an integrated cross section for $^{208}Pb+^{208}Pb \to AA'+kn$ as a
function of numbers of emitted neutrons (k) for different, somewhat arbitrary, conditions
on high photon energies. In panel a) we discuss how the cross section depends on the
lowest point where $P_k$ are set to constants. In this case the upper integration limit
is set to $E_{\gamma}^{max}$ = 2$\cdot$10$^8$~MeV.
In panel b) we test the dependence of the cross sections on the upper integration limit
$E_{\gamma}^{max}$. In this case the constant values of $P_k$ are set to those for 
$P_k(E_{\gamma}  = 200$~MeV) calculated in our TCM. 
The excitation energy of $^{208}$Pb for $E_{\gamma}  = 200$~MeV is around 100~MeV. In Ref.~\cite{fabris} the fission probability for nearby $^{200}$Pb at $E_{exc}\approx100$~MeV is almost 30\% thus the evaporation channel is damped. Moreover, testing various energy cut in our approach these values gave the lowest $\chi^2$ values when comparing to the ALICE data \cite{ALICE502a}.

The cross sections for $^{208}Pb+^{208}Pb  \to AA'+kn$ potentially strongly
depend on the high energy ($E_{\gamma} $) behavior of the $\gamma+ ^{208}Pb \to kn+A'$ cross section.
A good description of the ALICE data can be obtained assuming that the region of high-$E_\gamma$
($E_\gamma <$  100 GeV) plays a crucial role in production of up to five neutrons. In our opinion this requires
further model studies in the future.

\begin{figure}[h]
\includegraphics[width=8.9cm]{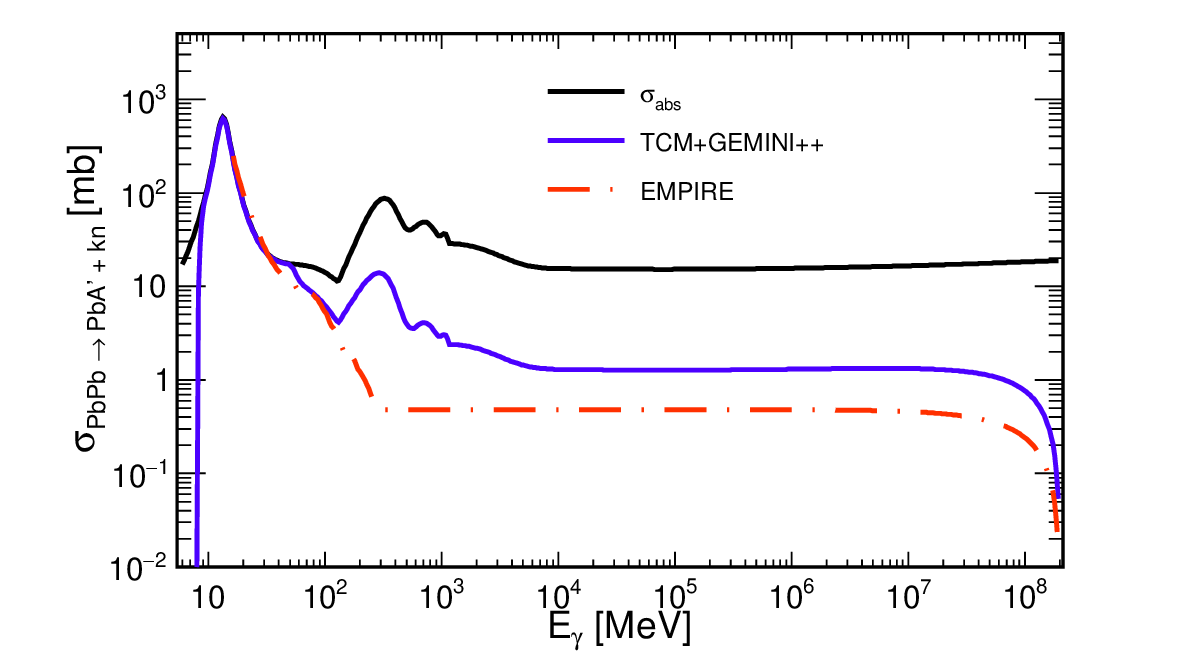}
\caption{The sum of cross sections for emitting up to 5 neutrons for TCM+GEMINI++ (blue solid line), EMPIRE (red dash-dot line) and total absorption cross section (black solid line) \cite{cs1, LEPRETRE1981, cs2, cs3, cs4, cs5}.
}
\label{fig:delta_reso}
\end{figure}

In Fig.~\ref{fig:delta_reso} we show consequences of the assumption $P_k = const$
for the cross section $\sum^{5}_{k=1} \sigma_{\gamma+ ^{208}Pb\to kn+A'}(E_\gamma)$ in
a very broad range of photon energy. It leads to about 10\% of the photoproduction
cross section shown for the comparison as the black solid line, up to extremely large
photon energies. It is impossible to verify at present how reliable is as no experimental
data exist. At high energies the result of our present approach (TCM) is very different than that
of EMPIRE.

The tests made above show the uncertainty of our integration, which gives less than 10\% of the error. There is also an inaccuracy connected with the number of steps in the integration procedure, but it gives less than 2\% for the highest energies. The statistical error is very small as GEMINI++ runs for $10^6$ events and HIPSE generates 5$\cdot10^3$ events.

\section{A new prescription for multiple
photon exchanges}

\noindent
In this section, we calculate cross section for a given number
of neutrons including the exchange of one and multiple photons.
We use a formalism discussed partly in \cite{reldis}.
We explicitly include up to 4 photon exchanges.

The simplest model (one photon exchange + equilibrium) is almost right 
for one and two neutron emissions (Table~\ref{tab:theory_ALICE}) but does not describe in detail 
the emission of 3, 4, 5 neutrons. In Table~\ref{tab:theory_ALICE} results for single photon exciting single nucleus are displayed and compared with experimental data from the ALICE Pb+Pb ultraperipheral collisions at energy $5.02$~TeV \cite{ALICE502a}. Pure GEMINI++ for k=3, 4, 5n gives the lowest estimation due to the missing contributions from high-energy tails in $E_\gamma$ spectra. The TCM improves the situation considerably.

\begin{table}[h]
\caption{Total cross sections (in barn) for a fixed number of neutron
  emission in UPC with energy $5.02$~TeV. By TCM we mean here Dirac
  delta + step-like probability function including GEMINI++. Pure GEMINI++ (GEM.), HIPSE+GEMINI++ (HG) and EMPIRE results are also presented. Here, only one photon exchange is taken under consideration (Fig.~\ref{fig:multipole_photon}a).
}
\centering
\begin{tabular}{c| r r r r r} 
\hline \quad kn \quad\quad&&&$\sigma$~[b]\quad\quad\quad\quad\quad&\\ 
\hline
   & \quad  GEM.  \quad & \quad  TCM  \quad &  EMPIRE &  \quad \quad \quad HG \quad \quad    \\ \hline
  1         &93.71   & 98.75    &  100.19   & 122.11     \\
 2          & 25.06   & 25.55  & 24.71      & 15.14  \\
 3          &  3.05   & 6.07   & 5.41       & 4.53  \\
 4          &  2.32   & 6.32   & 5.37       & 3.42  \\
 5          &  1.51   & 4.91   & 3.24       & 3.67  \\
\hline
\end{tabular}
\label{tab:theory_ALICE}
\end{table}

\begin{figure}[!h]
 \centering
 \setlength{\unitlength}{0.1\textwidth}
\begin{picture}(5,3)
\put(0.2,1.5){\includegraphics[width=4cm]{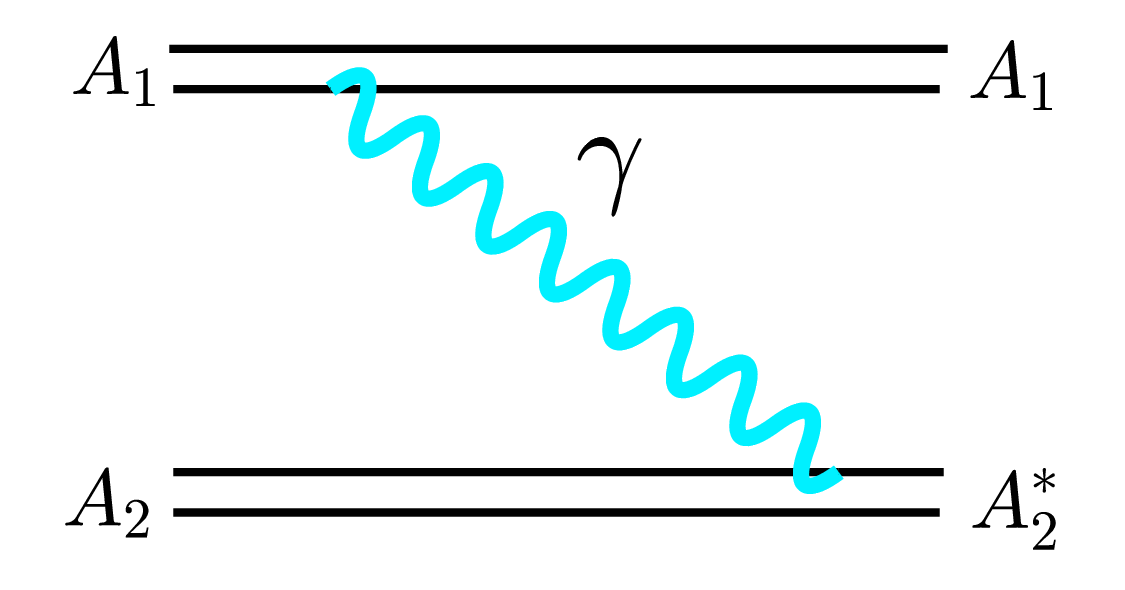}}
\put(0.2,2.1){a)} 
    \put(2.5,1.5){\includegraphics[width=4cm]{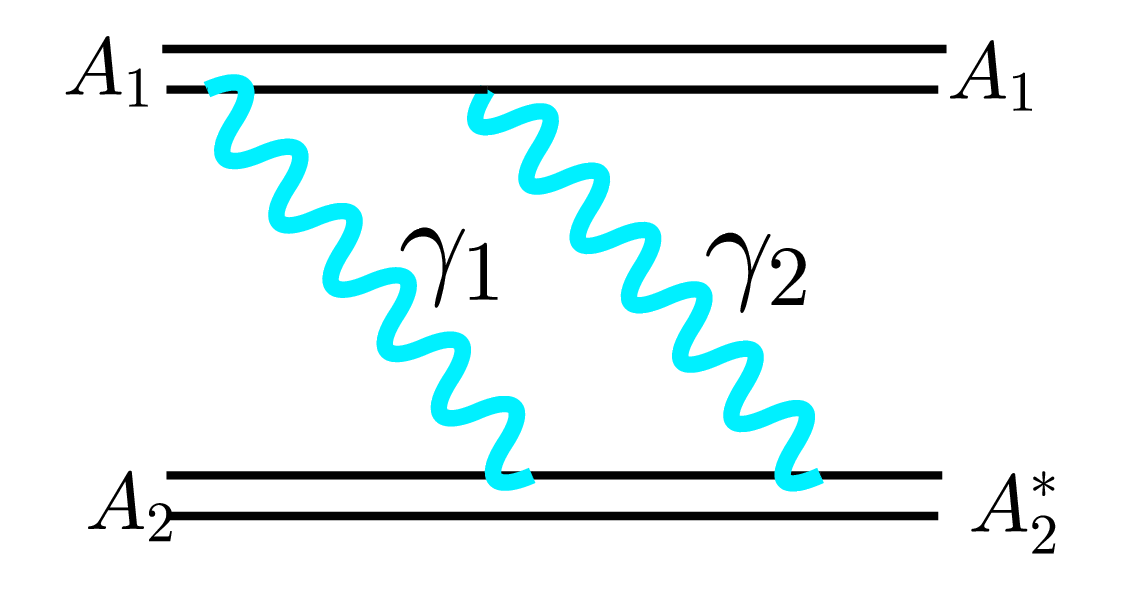}}
\put(2.5,2.1){b)}  
    \put(2.5,0){\includegraphics[width=4cm]{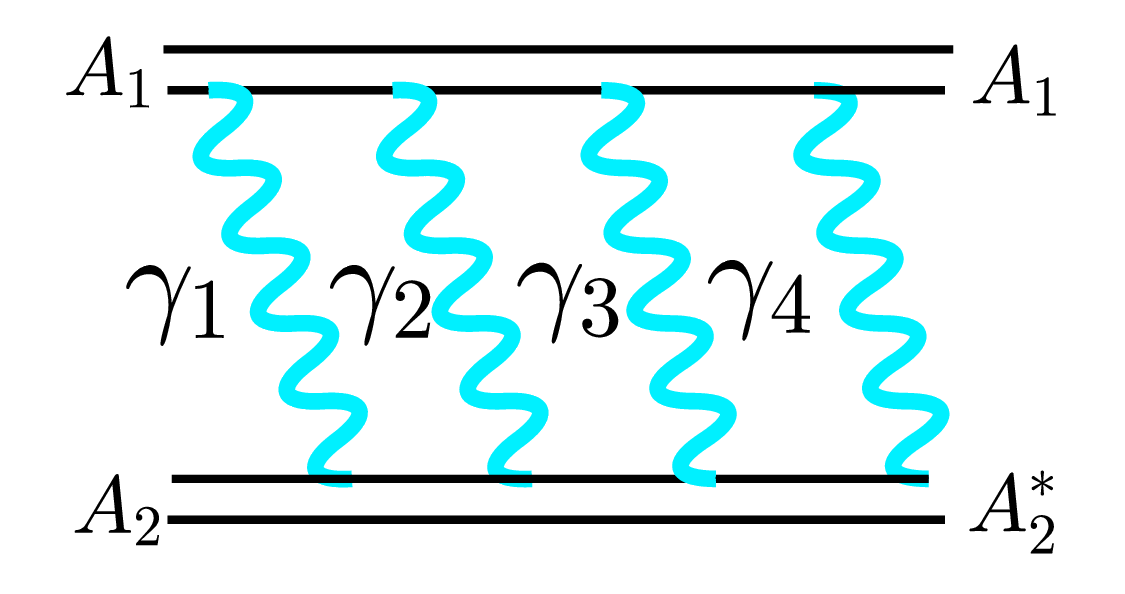}}
\put(0.2,0.6){c)}
     \put(0.2,0){\includegraphics[width=4cm]{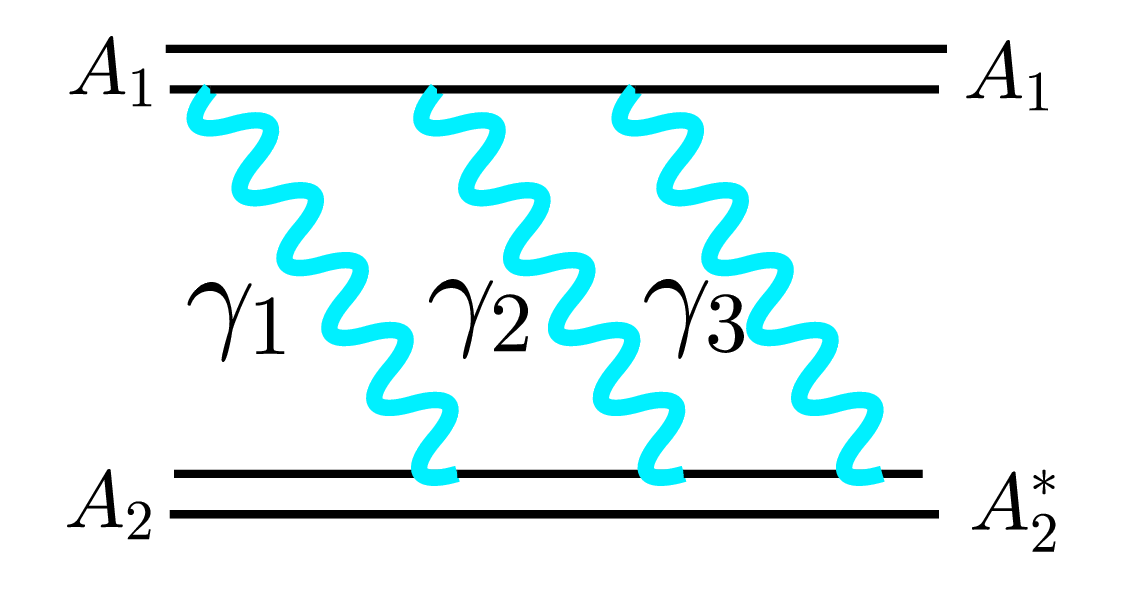}}
\put(2.5,0.6){d)} \end{picture}
\caption{Photon absorption by single nucleus ($A_2$). In panel a) one-photon absorption (LO), b) two-photon absorption (NLO), c) three-photon absorption (NLO$_2$) d) four-photon absorption (NLO$_3$).}
\label{fig:multipole_photon}
\end{figure}
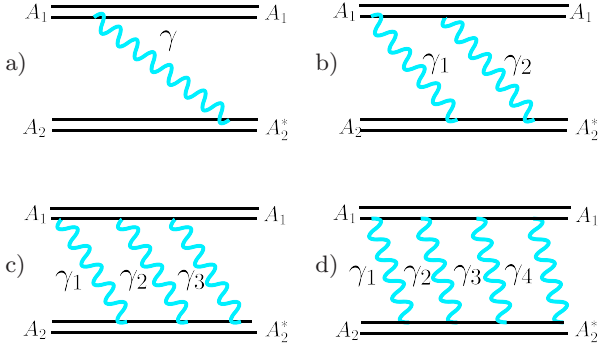

Thus we check to which extent multiple photon exchanges, as illustrated
in Fig.~\ref{fig:multipole_photon} can be responsible for the clear 
disagreement with the ALICE data for the number of emitted neutrons 
$n >2$.
Each additional photon exchange leads, on average, to higher nucleus excitation and, as a consequence, the emission of a bigger number of
neutrons.
As the 1-photon excitation is the most probable, let's call this process: "leading order" (LO), 2-photon gives smaller correction thus it will be "next-to-leading order" (NLO), 3-photon absorption- "next-to-next-to-leading order" (NLO$_2$), 4-photon absorption- "next-to-next-to-next-to-leading order" (NLO$_3$), what is shown in Fig.~\ref{fig:multipole_photon}, respectively.

The cross section for $k$ neutron emission from one nucleus in 
$A A \to A^{*} A$ collisions can be formally written as
\begin{equation}
\frac{d \sigma^k}{d E_{\gamma}} = \sum_{i=1}^4  
\frac{d \sigma_i^k}{d E_{\gamma,i}},
\label{sum}
\end{equation}
%
where $i$ is the number of photons exchanged
(we numerically include up to 4 photons) and
\begin{equation}
\frac{d \sigma_i^k}{d E_{\gamma,i}} = \frac{d \sigma_i}{d E_{\gamma,i}}
P_k(E_{exc}) \;,
\end{equation}
where $P_k(E_{exc})$ is a relative probability to
produce $k$ neutrons at excitation energy $E_{exc}=\sum_{i=1}^4 E_{\gamma,i}$.
This probability density can be calculated e.g. in the Hauser-Feshbach 
approach (see Ref.\cite{Klusek-Gawenda:2013ema}). The$\frac{d \sigma_i}{d E_{exc}}$ contains implicitly weight factors, 
$W_i(b) \approx \frac{exp(-m(b))}{i!}$ due to indistinguishable exchange photons to avoid double counting
(see, e.g.,\cite{reldis}).

The cross section for $i$ photon absorption by single nucleus in $A A \to A A^*$ collision 
can be deduced from Eq.~(\ref{eq:sigma_tot_neutron_sum}) and (\ref{one-photon_approximation}):
\begin{equation}
\begin{split}
    \frac{d \sigma_i}{d E} =  \int d E_1 ... d E_i \delta(E_1 + ... + E_i-E)
N(E_1,b) ... N(E_i,b)\\
\sigma_{\gamma+ A \to A^\prime}(E_1) ... \sigma_{\gamma+ A \to A^\prime}(E_i) W_i(b)
d^2 b \; .
\label{eq:singlenucleus}
\end{split}
\end{equation}

%

In Table \ref{tab:gemini_n_gamma}, we present our result for the cross section
for a given number of photon exchanges.
The multiphoton exchange processes were studied before in \cite{baur}, where the independence of the photon emission was discussed. The high energy of the colliding nuclei outbalances the photon energy. In consequence different interactions can be treated independently. The possible effect of correlation of neutrons emitted after mutual GDR excitation discussed e.g. in Baur et al. \cite{baur} cannot be observed in high energy collider mode when particles are emitted in very forward directions due to kinematical boost.

Here, for each 
photon exchange ($i$), we assume $\delta E_{exc}^{(i)} = E_{\gamma}^{(i)}$ where $\delta E_{exc}$ is a part
of the excitation due to individual photon exchange.
We observe an improvement, e.g. for three neutron emission the model of two-photon exchange enhances the cross section 
by 10\%, but still something is missing.
\footnote{Are mutual excitations (simultaneous excitation of 
both collision partners) important in this context?
Two questions are related to this issue.
Are mutual excitations eliminated by the way the experimental
data of the ALICE collaboration is obtained?
Does ALICE require no neutron emission from the second nucleus?}

\begin{table}
\caption{Cross section for emission of $k$ neutrons
via exchange of $i =$ 1, 2, 3 and 4 photons for $\sqrt{s_{NN}}=5.02$~TeV. $E_{exc}=E_{\gamma}$ approximation and GEMINI++ code are used here. 
}
\begin{tabular}{c r r r r r r}
\hline kn &&&$\sigma$~[b]\quad\quad&\\ 
\hline
   & LO & NLO  & NLO$_{2}$  &NLO$_{3}$  & all \\
\hline
1 $\quad$  &  $\quad$93.713	&	$\quad$0.008	& $\quad$0.000&	   $\quad$0.0000 & $\quad$93.721    \\
2 $\quad$  &         25.057	&	       0.429	&	 0.001    &		  0.0000     &        25.487   \\
3 $\quad$  &         3.054	    &	       0.385	&	 0.023    &		  0.0000     &        3.462    \\
4 $\quad$  &         2.316	    &	       0.103	&	 0.080    &		  0.0006     &        2.499    \\
5 $\quad$  &         1.505	    &	       0.038	&	 0.027    &		  0.0229     &        1.593    \\
\hline
\end{tabular}\label{tab:gemini_n_gamma}
\caption{Cross section for emission of $k$ neutrons
via exchange of 1, 2, 3 and 4 photons for $\sqrt{s_{NN}}=5.02$~TeV. Two-component
model TCM ($E_{exc}<E_\gamma$) is used for each exchange as described in the text.
}
\begin{tabular}{c r r r r r r}
\hline kn &&&$\sigma$~[b]\quad\quad&\\ 
\hline
     & LO & NLO  & NLO$_{2}$  &NLO$_{3}$  & all \\
\hline
1 $\quad$  &$\quad$98.749 &$\quad$0.295   & $\quad$ 0.067      &$\quad$0.0010&$\quad$ 99.112    \\
2 $\quad$  &   25.549 &	    0.637   &	    0.095     &		  0.0014 &	 26.282	   \\
3 $\quad$  &   6.065  &	    0.410   &	    0.084     &		  0.0011 &	 6.560 	   \\
4 $\quad$  &   6.324  &	    0.260   &	    0.121     &		  0.0015 &	 6.706 	   \\
5 $\quad$  &   4.909  &	    0.191   &	    0.076     &		  0.0013 &	 5.178 	   \\
\hline
\end{tabular}\label{tab:tcm_n_gamma}
\end{table}

The one photon exchange has a maximum for $1 n$ production,
the two-photon exchange for $3 n$ production, the three
photon exchange for $4 n$ production.
Moreover Fig.~\ref{fig:UPC_hipse} shows distribution of excitation energy
for a given number of neutrons and exchanged photons
which presents our point even better.

In Table \ref{tab:tcm_n_gamma}, we present our result for the different numbers
of photon exchanges for the two-component model (TCM)
to simulate pre-equilibrium effects. We get the higher-order corrections of the order of 10\% for k=3, 4, 5.

\begin{figure}[h]
 \centering
 \setlength{\unitlength}{0.1\textwidth}
\begin{picture}(5,10)
\put(0,7.5){\includegraphics[width=8cm]{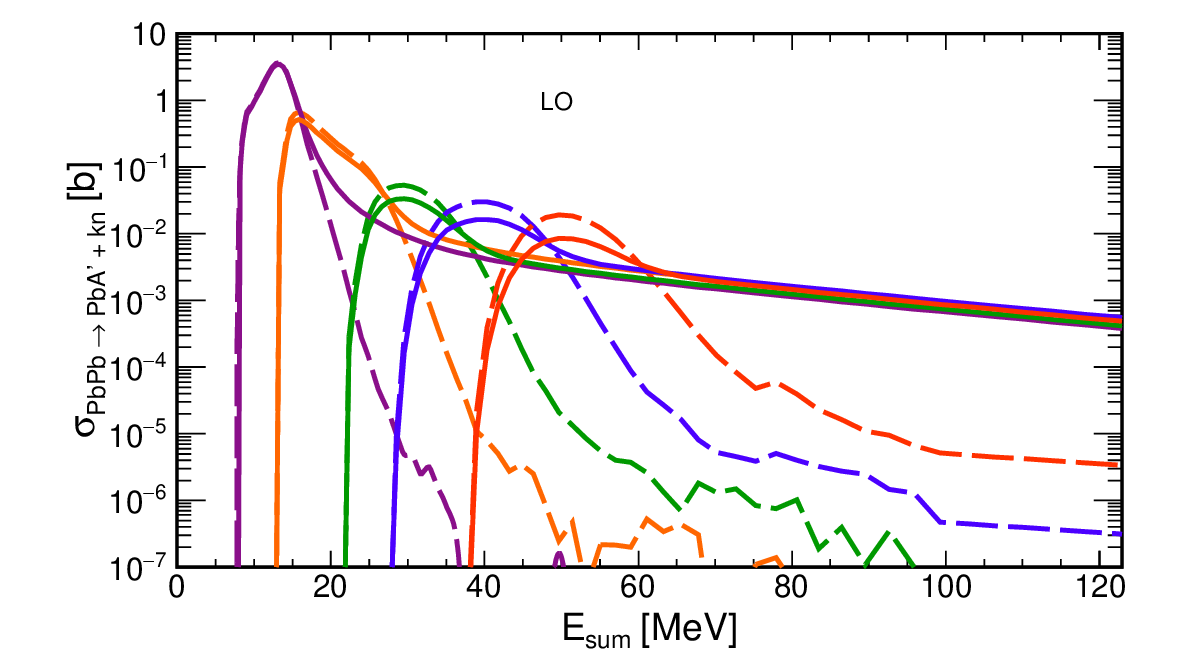}}
\put(3.9,9.6){a)} 
    \put(0,5){\includegraphics[width=8cm]{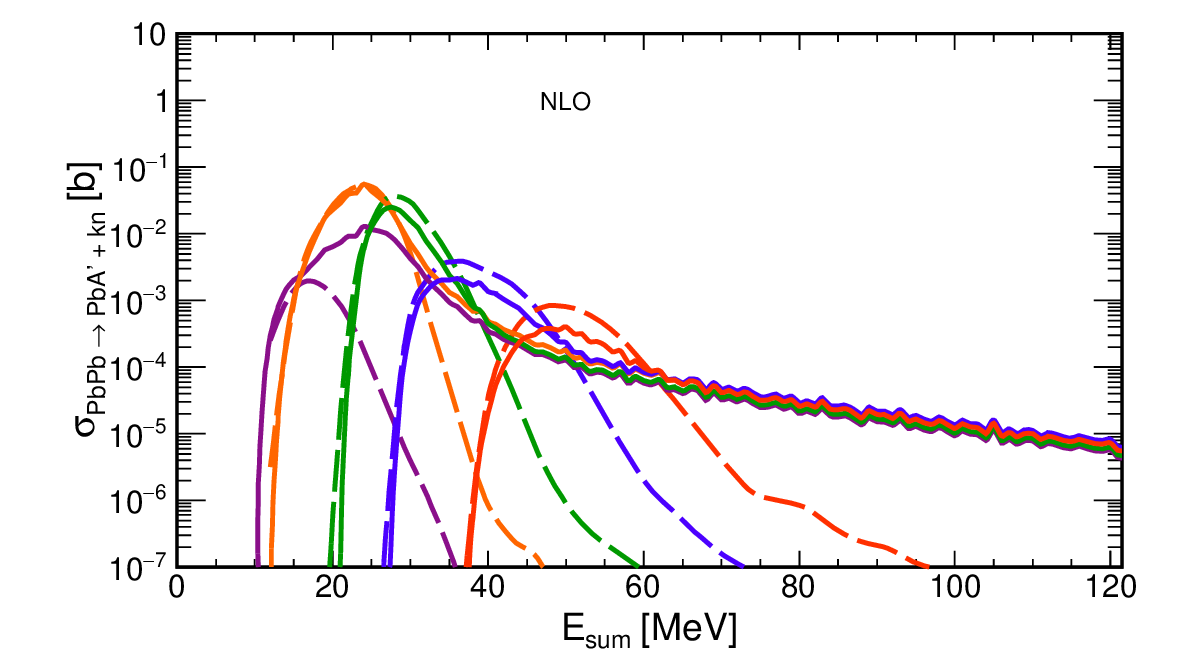}}
\put(3.9,7.1){b)}  
    \put(0,2.5){\includegraphics[width=8cm]{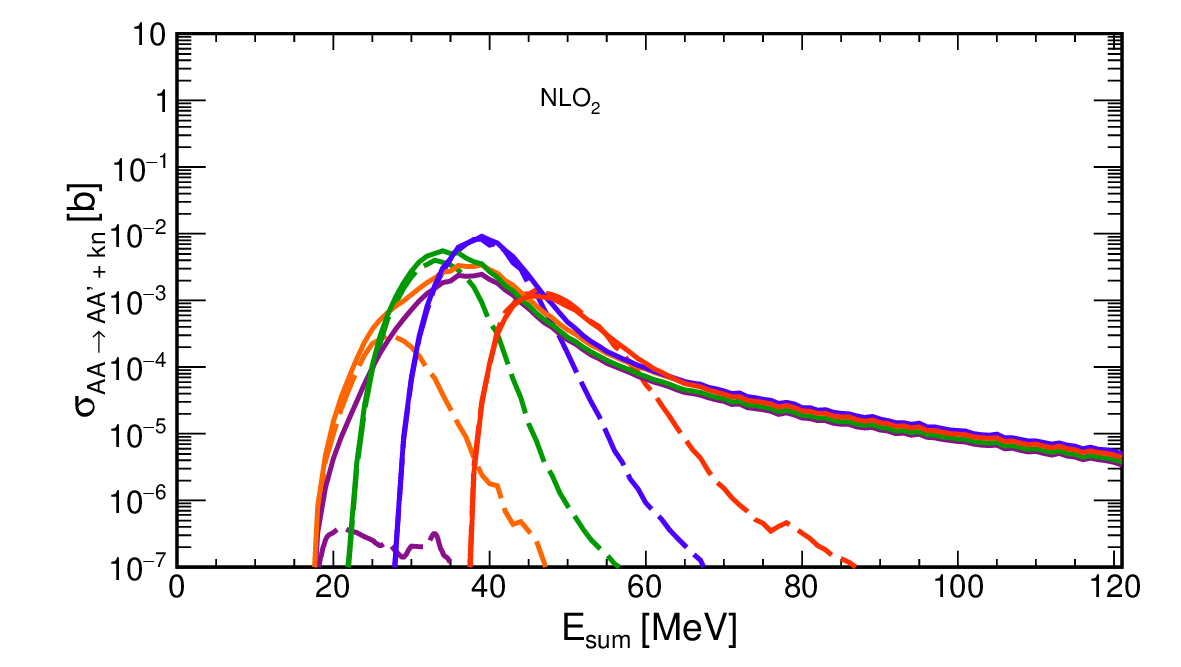}}
\put(3.9,4.6){c)}
     \put(0,0){\includegraphics[width=8cm]{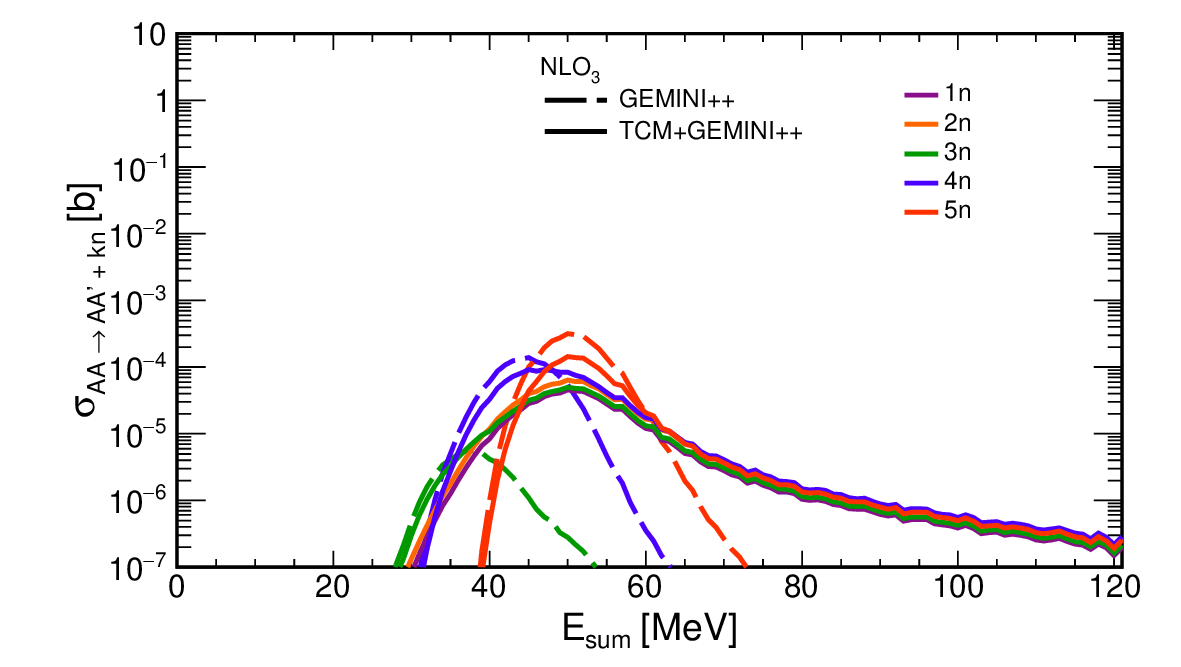}}
\put(3.9,2.1){d)} \end{picture}
\caption{
Distribution in excitation energy for a given number of exchanged 
photons (a) 1$\gamma$-rays, (b) 2$\gamma$-rays,(c) 3$\gamma$-rays,(d) 4$\gamma$-rays and a given number of emitted neutrons kn= 1, 2, 3, 4, 5. 
}
\label{fig:k_photons}
\end{figure}
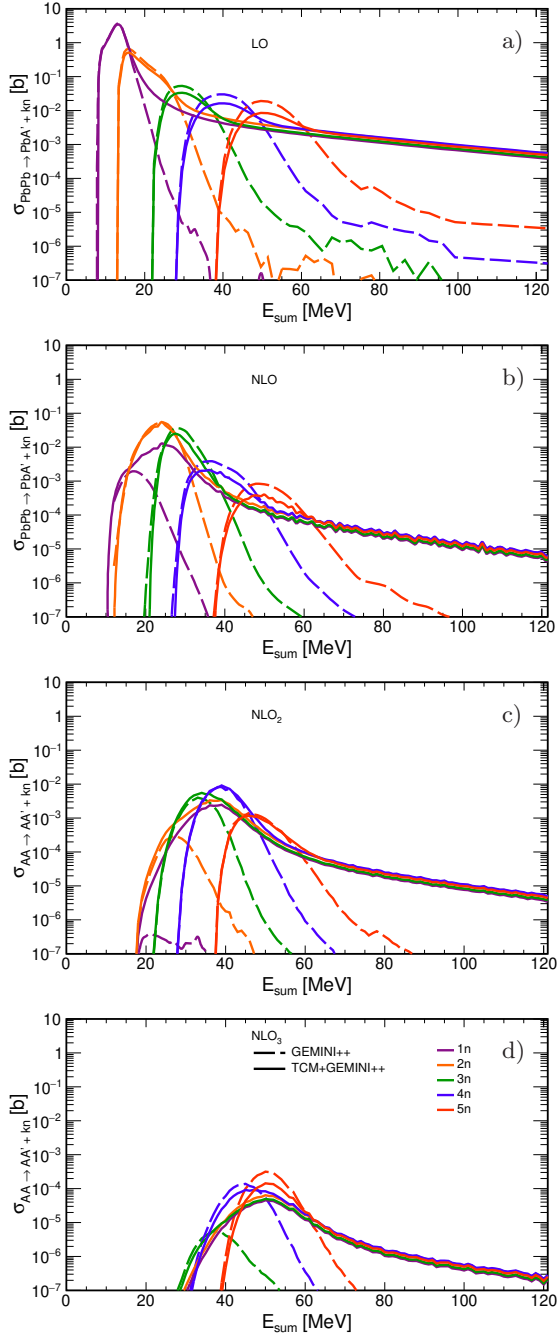

For completeness in Fig.~\ref{fig:k_photons} we present
distribution in excitation energy, estimated as a sum of photon energies that hit the single nucleus, for a given number of photons
and neutrons using TCM as described in the text above.\\

\section{Mutual exchange of photons between nuclei}
\begin{figure}[!h]
 \centering
 \setlength{\unitlength}{0.1\textwidth}
\begin{picture}(5,3)
\put(0.2,1.5){\includegraphics[width=4cm]{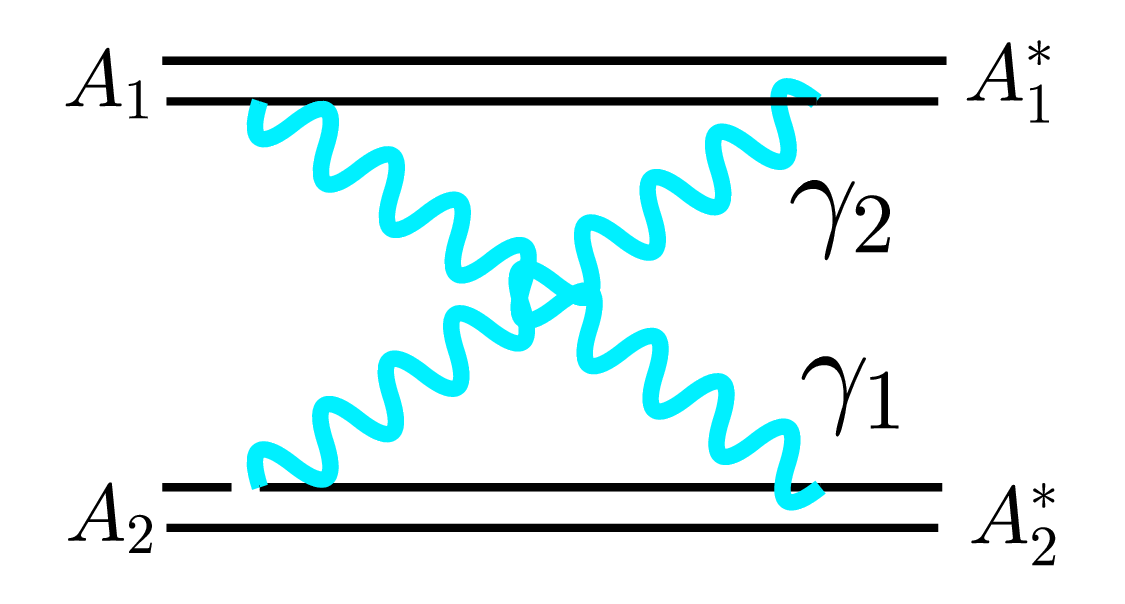}}
\put(0.2,2.1){a)} 
    \put(2.5,1.5){\includegraphics[width=4cm]{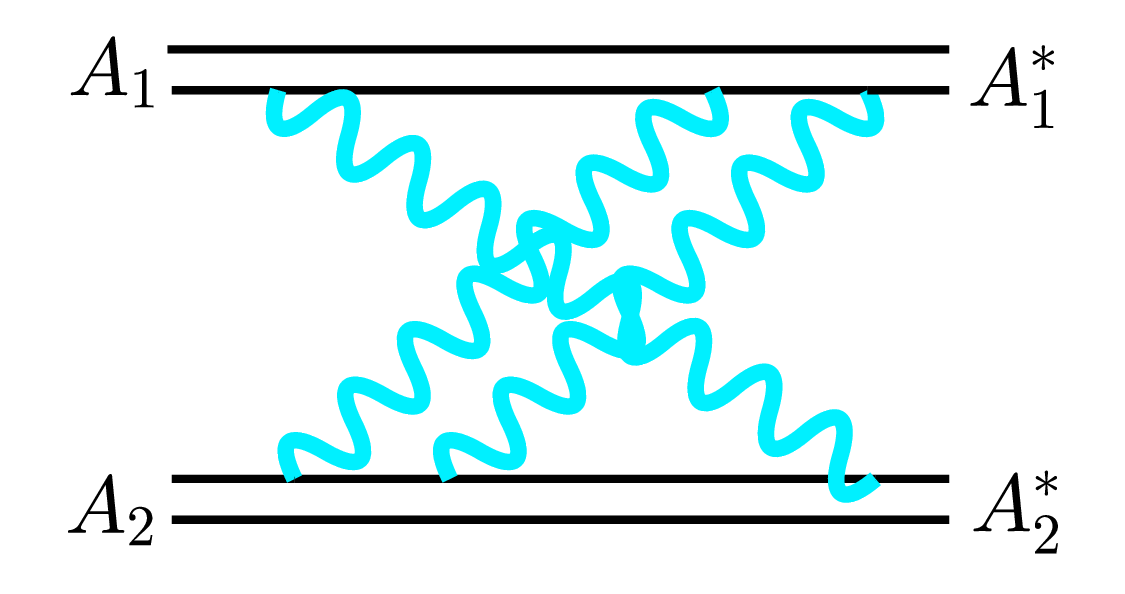}}
\put(2.5,2.1){b)}  
    \put(2.5,0){\includegraphics[width=4cm]{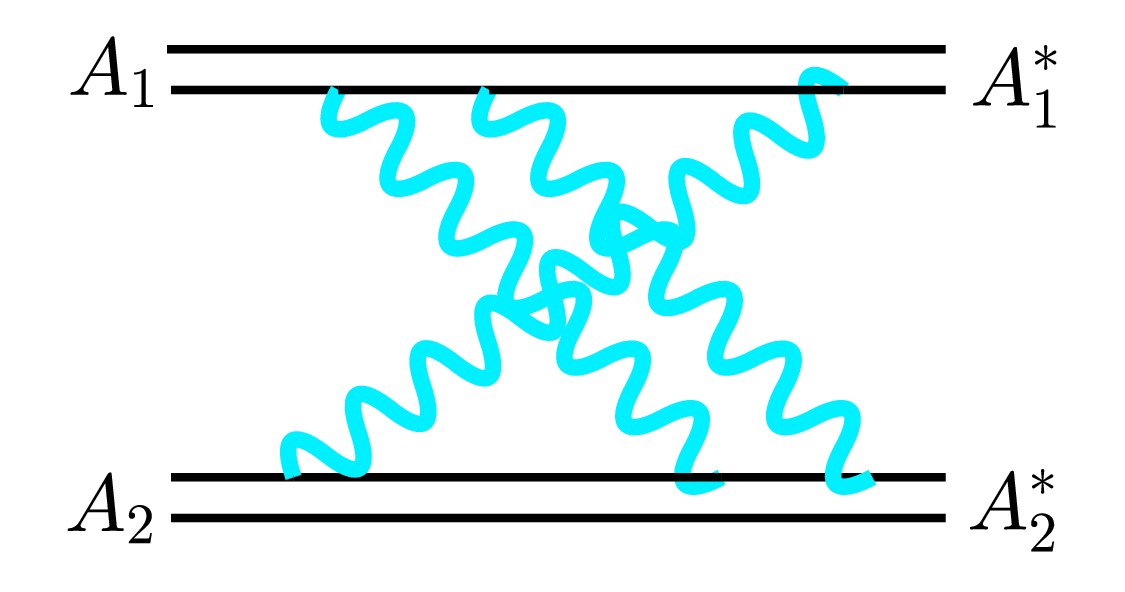}}
\put(0.2,0.6){c)}
     \put(0.2,0){\includegraphics[width=4cm]{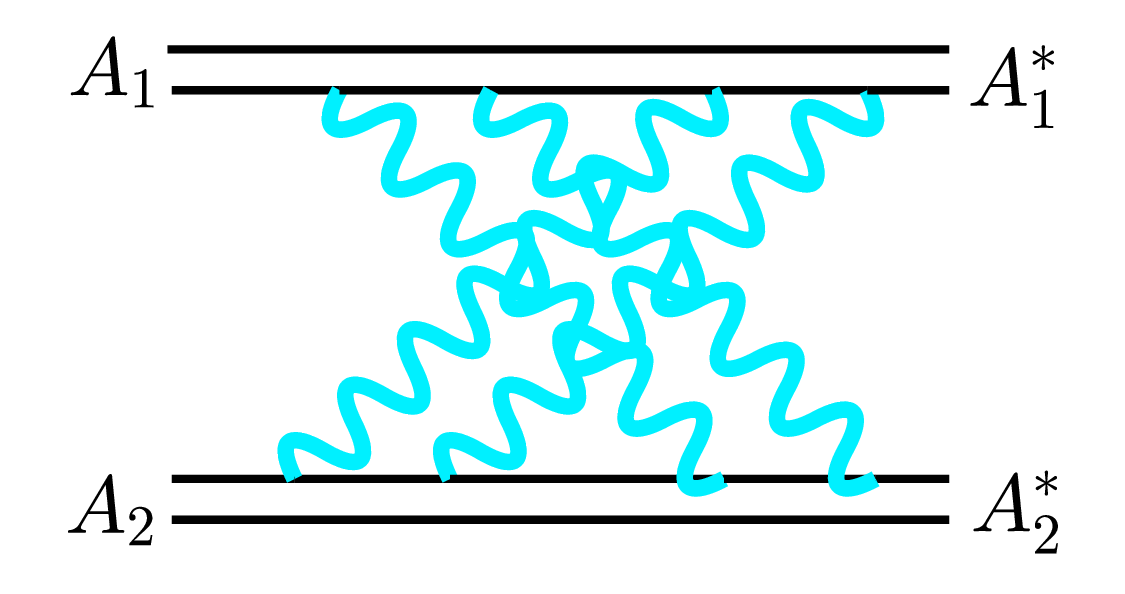}}
\put(2.5,0.6){d)} \end{picture}
\caption{Multiple photon exchanges - simultaneous
excitation of both nuclei, called mutual excitation.}
\label{fig:mutual_excitations}
\end{figure}
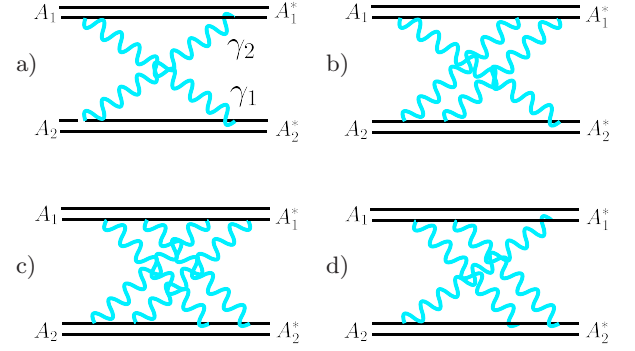

Also mutual excitations, understood as simultaneous excitations of 
both nuclei must be, in principle, included.
Some examples of corresponding photon exchanges are shown in
Fig.~\ref{fig:mutual_excitations}.

The corresponding leading-order cross section (diagram (a)) for 
the mutual excitation, following Eq.~(\ref{eq:singlenucleus}), (see e.g.~\cite{reldis}) reads:
\begin{equation}
\begin{split}
\sigma_{A_1A_2 \to A_1^*A_2^*}(E_1,E_2) =
\int d E_1 d E_2 d^2b \\
N(b,E_1) \sigma_{\gamma+A_2 \to A_2^*}(E_1)
N(b,E_2) \sigma_{\gamma + A_1\to A_1^*}(E_2) \; .
\end{split}
\label{LO_mutual}
\end{equation} 
For example, the cross section for $1$ neutron emission from the first excited nucleus $A_1^*$ and $1$ neutron emission from the second excited nucleus $A_2^*$ can be then written as
\begin{eqnarray}
\frac{d\sigma_{AA\to A_1^*A_2^*}}{dE_1dE_2} &=& \int d^2 b N(b,E_1) P_{1kn}^{1\gamma}(E_1) \sigma_{\gamma +A_1  \to A_1^* }(E_1) \nonumber \\
&\times&  N(b,E_2) P_{2kn}^{1\gamma}(E_2) \sigma_{\gamma +A_2  \to A_2^* }(E_2)   ,
\end{eqnarray}
where $P_1(E_1)$ and $P_1(E_2)$ are familiar one neutron emission 
probabilities which have to be modeled, e.g. in the Hauser-Feshbach
approach or in TCM.

One can also consider $(n_{\gamma_1}, n_{\gamma_2})$ = (1,2) + (2,1) + (2,2) +...
cross sections (see extra columns in Table~\ref{tab:neutrons_mutual_excitation}) which only slightly improve
the agreement with the ALICE data. Here $n_{\gamma_1}$ is a number
of photons emitted by the first and absorbed by the second nucleus.

The cross section for excitation of the first nucleus in mutual interactions reads:

\begin{eqnarray}
\frac{d\sigma_k}{dE_{A_1}} = \sum^4_l \sum_{i,j} \int dE_{A_2} \frac{ d\sigma_{k,l}^{(i,j)} }{ dE_{A_1}^{(i)}dE_{A_2}^{(j)} },
\end{eqnarray}

where:

\begin{eqnarray}
E^{(i)}_{A_1} = \sum^{i}_{k=1} E_{A_1,k}; \\
E^{(j)}_{A_2} = \sum^{j}_{l=1} E_{A_2,l}.
\end{eqnarray}

In general, mutual excitations with the same number of exchanged
photons are concentrated at smaller
impact parameter than single excitations which is due to a bigger
number of flux factors involved
(for leading order terms one flux factor for single excitation,
versus two flux factors for mutual excitation 
(see Eq.(\ref{LO_mutual}))).

The cross sections for mutual excitations are summarized in
Table~\ref{tab:neutrons_mutual_excitation}.
These numbers are much smaller than those for single nucleus excitation 
(see Tables~\ref{tab:theory_ALICE}, \ref{tab:gemini_n_gamma}, \ref{tab:tcm_n_gamma}).
So mutual excitations cannot explain the discrepancy with the ALICE data.
In the mutual excitation also the dominant role plays the exchange of single photons.

\begin{figure}[!bt]
\hspace{-0.7cm}
\includegraphics[width=9.2cm]{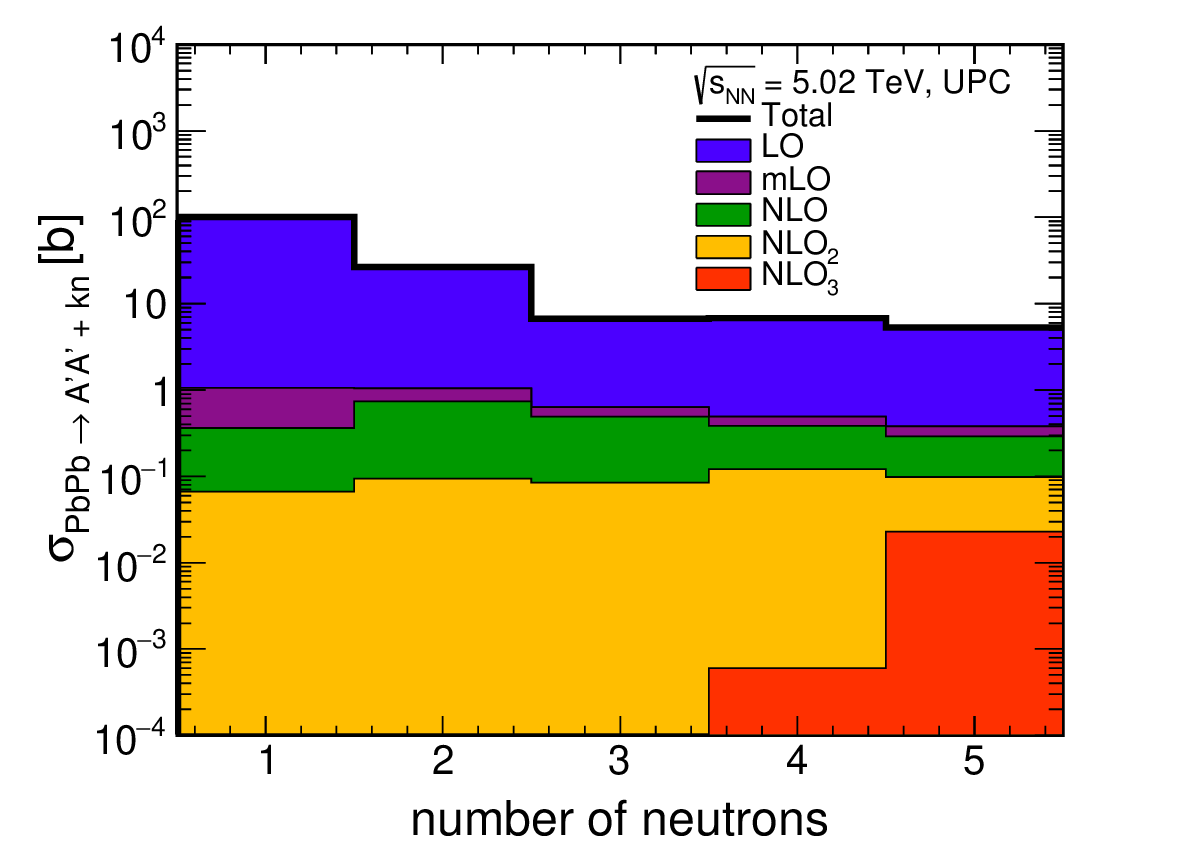}
\caption{Various components of the neutron multiplicities cross section from photon-induced Pb nuclei such as: single nucleus-one photon (LO), single nucleus and x+1 photons (NLOx) and both nuclei excited by photons exchange.}
\label{fig:mutual_n}
\end{figure}

A comparison of various components consisting of the final neutron emission cross section is presented in Fig.~\ref{fig:mutual_n}. Logarithmic y-axis can be misleading but allows to show small contributions from 3 and 4 photon exchanges. The biggest influence has the single photon-single nucleus excitation process and other cases are just a correction. However, interesting is the fact that they are almost independent of the number of emitted neutrons.

\begin{figure}[!bt]
    \centering
        \includegraphics[width=0.48\textwidth]{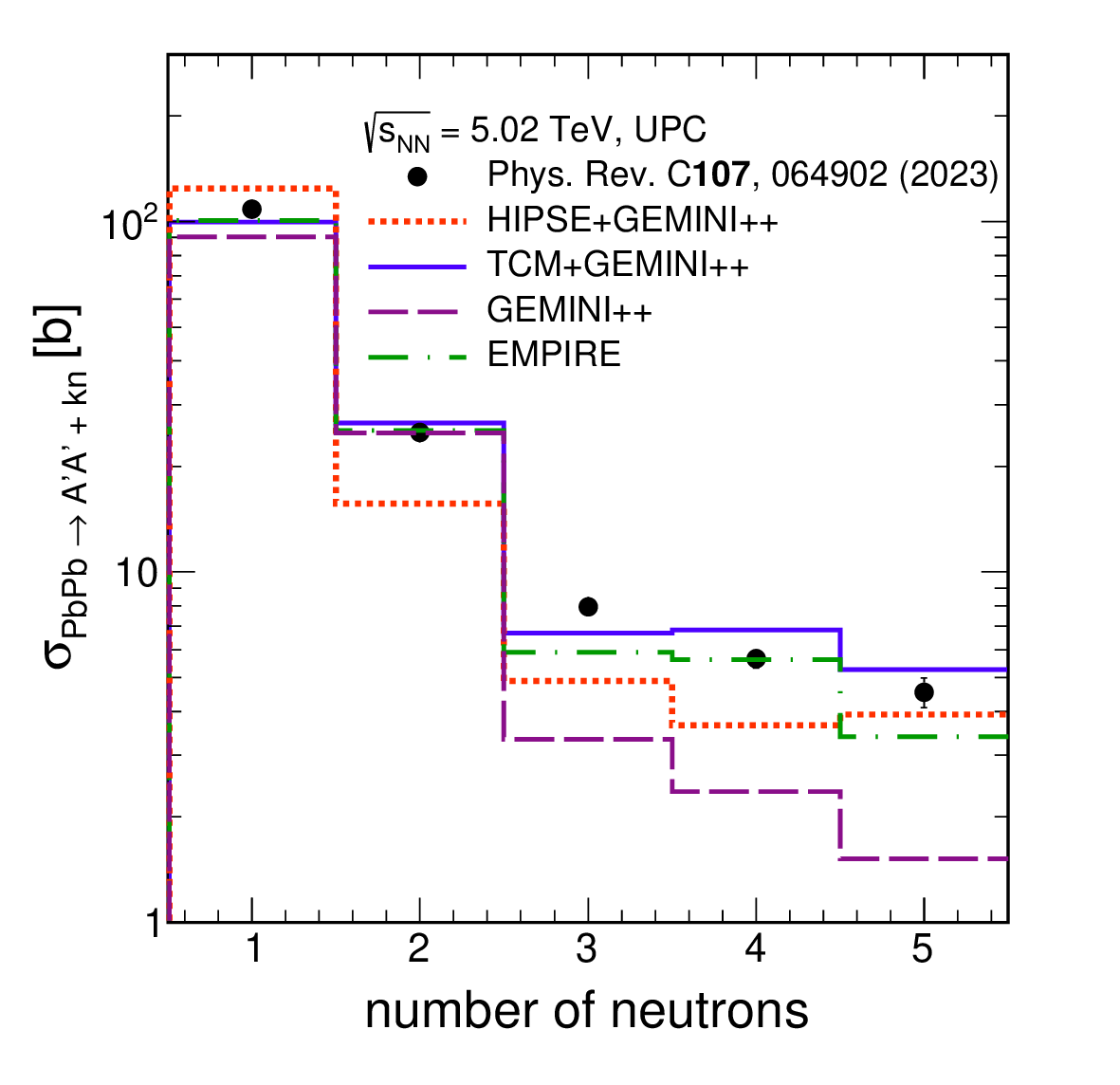}
    \caption{Total cross section of emission of a given number of neutrons in lead-lead UPC, with collision energy $\sqrt{s_{NN}}$~=~5.02~TeV. The dots represent results from the ALICE experiment \cite{ALICE:2012aa}.}
    \label{fig:UPC_res}
\end{figure}

The Fig.~\ref{fig:UPC_res} summarizes the observation made in our paper. 
The effect of pre-equilibrium emission is illustrated by comparison between GEMINI++ and TCF+GEMINI++ or HIPSE+GEMINI++. For emission of higher multiplicity neutrons, important are the high energy photons. 

\begin{table}
\caption{Cross sections in barns for mutual excitations 
at $\sqrt{s_{NN}}=5.02$~TeV
for a given number of neutrons, assuming full equilibrium
in both excited nuclei, i.e. $P_k(E_1)$ and $P_k(E_2)$ are 
estimated by the TCM. 
} 
\begin{tabular}{ccccc}

\hline kn &&&$\sigma$~[b]\quad\quad\quad\quad&\\ 
\hline
 & \quad mNLO\quad  & \quad mNLO$_2$ \quad&  \quad mNLO$_2$ \quad & \quad mNLO$_3$ \quad  \\
\hline
$A_1$, $A_2$   &  $1\gamma-1\gamma$  &  $1\gamma-2\gamma$ &  $2\gamma-1\gamma$ &  $2\gamma-2\gamma$  \\
\hline 
\hline
1 \quad 1     & 0.2977  &0.0264 &0.0264 &0.0018 \\ 
1 \quad 2     & 0.0868  &0.0568 &0.0077 &0.0045 \\
1 \quad 3     & 0.0262  &0.0365 &0.0023 &0.0022 \\
1 \quad 4     & 0.0295  &0.0232 &0.0026 &0.0011 \\
1 \quad 5     & 0.0295  &0.0171 &0.0021 &0.0008 \\
\hline     	                	 
2  \quad  1   & 0.0868  &0.0077 &0.0568 &0.0045 \\
2  \quad  2   & 0.0253  &0.0166 &0.0166 &0.0111 \\
2  \quad  3   & 0.0077  &0.0107 &0.0051 &0.0056 \\
2  \quad  4   & 0.0086  &0.0068 &0.0057 &0.0030 \\
2  \quad  5   & 0.0070  &0.0050 &0.0046 &0.0021 \\
\hline     	              		 
3  \quad  1   & 0.0262  &0.0023 &0.0365 &0.0022 \\
3  \quad  2   & 0.0077  &0.0051 &0.0107 &0.0056 \\
3  \quad  3   & 0.0023  &0.0032 &0.0032 &0.0026 \\
3  \quad  4   & 0.0026  &0.0021 &0.0037 &0.0013 \\
3  \quad  5   & 0.0021  &0.0015 &0.0030 &0.0009 \\
\hline     	            		 
4 \quad   1   & 0.0295  &0.0026 &0.0232 &0.0011 \\
4 \quad   2   & 0.0086  &0.0057 &0.0068 &0.0030 \\
4 \quad   3   & 0.0026  &0.0037 &0.0021 &0.0013 \\
4 \quad   4   & 0.0029  &0.0023 &0.0023 &0.0006 \\
4  \quad  5   & 0.0024  &0.0017 &0.0019 &0.0004 \\
\hline	   	   			 
5 \quad   1   & 0.0239  &0.0021 &0.0171 &0.0008 \\
5 \quad   2   & 0.0070  &0.0046 &0.0050 &0.0021 \\
5 \quad   3   & 0.0021  &0.0030 &0.0015 &0.0009 \\
5  \quad  4   & 0.0024  &0.0019 &0.0017 &0.0004 \\
5 \quad   5   & 0.0019  &0.0014 &0.0014 &0.0003 \\
\hline
\end{tabular}
\label{tab:neutrons_mutual_excitation}
\end{table}

\section{Charged Particle Emission}

The zero degree calorimeter (ZDC) can be used to measure both forward 
neutrons and protons \cite{ZDC_protons}.
Recently the ALICE collaboration presented the first results of
the proton measurement in ZDCs \cite{ALICE_protons}. 
They measured inclusive proton cross section as well as production 
of selected 1pkn channels with 1 proton and 1, 2, 3 neutrons. 
The obtained cross section for 1 proton emission is 40.4 b. 
This can be compared to our results collected in Table~\ref{tab5}.

\begin{table}[!bt]
\caption{Total cross sections (in barn) for a charged particle emission in UPC $^{208}$Pb+$^{208}$Pb with collision energy $\sqrt{s_{NN}}$ = 5.02~TeV calculated with pure GEMINI++(GEM.), TCM+GEMINI++ (TCM), HIPSE + GEMINI++(HG) and EMPIRE(EMP.). The cross section for exclusive channels of 1p1n, 1p2n and 1p3n and also inclusive channels 1pXn, 1dXn and 1$\alpha$Xn are compared with ALICE data (exp.), Ref.~\cite{ALICE_protons}.
}
\centering
\begin{tabular}{l r r r r r r r r} 
\hline Model &\multicolumn{6}{}{} \quad\quad \quad\quad \quad\quad\quad$\sigma$[b]\\ 
\hline
  & 1p1n        &   1p2n  & 1p3n  & 1pXn   &1dXn      & 1$\alpha$Xn  \\ \hline
 GEM.               &0&0& 0       &        19.56   &       6.91 &        30.66  \\
TCM          &0.66& 0.92 & 0.72            &       16.72   &         11.42 &     15.47\\
HG        &9.86& 0.82&    0.93             & 28.48   & 64.02 &  42.38 \\
  EMP.                &2.43&2.81&    0.23     &         7.06   &         2.65 &         0.28\\
exp.\cite{ALICE_protons} &1.05$\pm$0.07&1.35$\pm$0.21&1.58$\pm$0.57&40.4$\pm$1.7&-&-\\\hline
\end{tabular}
\label{tab5}
\end{table}
%
In Table~\ref{tab5} we also show cross section for the emission of 1 proton correlated with various numbers of neutrons: k=1, 2, 3 estimated by pure GEMINI++, two-component model with GEMINI++, HIPSE+GEMINI++ and EMPIRE approaches.

In the case of simultaneous emission of protons and a few neutrons, we assumed combined probability to be $P_{ik}^{p,n}(E_{\gamma})=P_{i}^{p}(E_{\gamma})P_{k}^{n}(E_{\gamma})$.

From GEMINI++ we cannot obtain probability for emission of one proton and one, two or three neutrons due to low proton emission rate in energy range: (10~--~50)~MeV. Thus, the zeros for pure GEMINI++ calculation can be understood as follows: in the Hauser-Feshbach approach, protons are emitted for $E_{exc}>$ 40~MeV while 1, 2 and 3 neutron channel are opened for  $E_{exc}<$ 40~MeV (Fig.~\ref{fig:gammaA_knX_exp}) i.e. simultaneous emission is not possible. However, integration over all possible 1pXn channels gives meaningful values. 

The situation improves for TCM+GEMINI++, where the large tails in $E_{\gamma}$ for neutron emission probability $P^n_k(E_{\gamma})$ appear.

In general, the EMPIRE platform gives the smallest cross section and the HIPSE+GEMINI++ shows the dominant role of the deuterons in the pre-equilibrium emission. 

With the present experimental setting of ZDC, other charged particles than protons are not measured by the proton ZDC.
This requires dedicated simulations, including details of the experimental apparatus, which goes beyond the scope of the present paper.

Our numbers seem to be smaller than those measured by the ALICE 
collaboration.
The details depend, however, on the model used. The further theoretical and experimental studies are necessary as the discrepancy between different models is quite significant. We get a reasonable agreement with individual $p, n$ channels.

\section{Conclusion}

In this paper, we have explored the mechanisms of neutron
production in ultrarelativistic, ultraperipheral 
$^{208}$Pb + $^{208}$Pb collisions.
In our approach, the Coulomb excitation is calculated as a convolution
of photon flux and photoabsorption cross section. The photon flux
is calculated in the equivalent photon approximation with approximate
or realistic photon fluxes. By realistic, we mean photon flux obtained
as a Fourier transform of the realistic charge distribution
in  $^{208}$Pb.
The nucleus is excited by the absorption of the exchanged photon.
In the previous paper of our group, we assumed $E_{exc} = E_{\gamma}$.
There only a small number of emitted neutrons (up to three) was 
considered.

Recently, the ALICE collaboration measured up to five neutrons emitted
from one nucleus at center-of-mass collision energy 
$\sqrt{s_{NN}}$ = 5.02~TeV.
The ALICE collaboration presented cross sections
for 1, 2, 3, 4 and 5 neutron production separately.
In our approach we have to use a model of emission of particles
from the Coulomb excited nucleus. Since typical excitation
energies are less than 150~MeV it is natural to assume 
the Hauser-Feshbach cascade emission as implemented in GEMINI++
program.

We have shown that the sketched above approach is able
to describe the larger neutron multiplicity ALICE data.
We have obtained slightly smaller cross sections than the measured ones for neutron multiplicities $M_n >$ 2.
This has forced us to reconsider some simplifications made in our previous
paper and at the initial stage of the present paper.  

The approach we have discussed here relies on allowing that
not the whole $E_{\gamma}$ energy goes to the statistically equilibrated
system. Still a part of an energy escapes in the form of pre-equilibrium emissions,
which means that $E_{exc}$ of the equilibrated nuclear system is smaller 
than $E_{\gamma}$ in the convolution approach. In this respect, we have 
proposed a two-component model. One component is treated as 
previously (full equilibration) but the second component assumes in a hidden way 
pre-equilibrium emissions. The relative probabilities of each component
may naturally depend on $E_{\gamma}$.
Such an approach changes cross sections for different
neutron multiplicities especially for k=3, 4, 5. We have also considered other approaches, such as HIPSE or EMPIRE. 

We have discussed the potential role of the high energy photons ($E_{\gamma}>140$~MeV) $\gamma+A\rightarrow kn+X$ for small neutron multiplicities relevant for this paper. This issue requires further model studies in the future.

In the first stage, we assumed a single photon exchange.
Next, we also included multiple photon exchanges. The latter are
more important at larger nuclear excitation, i.e. a bigger number
of emitted neutrons. We have reported some improvements.

There can be some other effects which may be responsible for
the persisting small disagreement. For example, the measured data may not be
fully UPC type; some peripheral collisions may play a difficult to quantify role. 
Secondly, we have considered simplified formulae for
multiple photons exchanges. However, a more refined treatment goes
beyond the scope of this paper.

Finally, we have also presented our results for the emission of p, d and $\alpha$-particles in different nuclear models used in the present studies. At the final stage of this paper new ALICE results for emission of protons became available. Thus we have confronted our estimations with experimental data also for simultaneous emission of protons and neutrons and found fair agreement.

\vspace{1cm}
{\bf Acknowledgments}

We are indebted to Igor Pshenichnov, Krzysztof Pysz, 
Sergei Ostapchenko, Tanquy Pierog, Wolfgang Sch\"afer,
Daniel Tapia Takaki and Chiara Oppedisano for discussion of topics related to the present paper.  This work was partially
supported by the Polish National Science Center Grant DEC2021/42/E/ST2/00350.

\bibliography{biblio}

\end{document}